\documentclass[%
 reprint,
 amsmath,amssymb,
 aps,
]{revtex4-2}
\usepackage[caption = false]{subfig}
\usepackage{graphicx}
\usepackage{dcolumn}
\usepackage{bm}
\usepackage{hyperref}
\usepackage{amsmath}
\usepackage[utf8]{inputenc} 
\usepackage[T1]{fontenc}
\def\apj{ApJ}
\def\apjl{ApJL}

\def\mnras{MNRAS}

\def\prd{PRD}
\def\jcap{JCAP}

\begin{document}

\preprint{APS/123-QED}

\title{Simulations of Axion Minihalos}

\author{Huangyu Xiao$^1$}
\author{Ian Williams$^2$}
\author{Matthew McQuinn$^2$}
\affiliation{$^1$Department of Physics, University of Washington,  Seattle, WA 98195,USA}
\affiliation{$^2$Department of Astronomy, University of Washington,  Seattle, WA 98195,USA}
\date{\today}
\begin{abstract}
The axion, motivated as a solution to the strong CP problem, is also a viable dark matter candidate. The axion field takes random values in causally disconnected regions if the symmetry breaking that establishes the particle occurs after inflation, leading to white-noise density fluctuations at low wavenumbers and forming dense minihalos with sub-planetary masses subsequently. There have been two recent proposals that appear capable of testing this scenario, namely using pulsar timing arrays and studying cosmological microlensing caustics.  Motivated by these proposals, we use N-body simulations to study the formation of substructures from white-noise density fluctuations. The density profiles of our relaxed axion minihalos can be described by the Navarro-Frenk-White profile, and the minihalos' concentration number agrees well with a simple, physically-motivated model. We develop a semi-analytic formula to fit the mass function from our simulation, which agrees broadly at different redshifts and only differs at factor of two level from classic halo mass functions. This analytic mass function allows us to consider uncertainties in the post-inflation axion scenario, as well as extrapolate our high-redshift simulations results to the present. 

Our work estimates the present-day abundance of axion substructures, as is necessary for predicting their effect on cosmological microlensing caustics and pulsar timing. Our calculations suggest that if pulsar timing and microlensing probes can reach recent sensitivity forecasts, they may be sensitive to the post-inflation axion dark matter scenario, even when accounting for uncertainties pertaining to axion strings.  For pulsar timing, the most significant caveat is whether axion minihalos are disrupted by stars, which our estimates show is mildly important at the most relevant masses. Finally, as our gravitational simulations are scale invariant, the results can be extended to models where the dark matter is comprised of other axion-like particles and even clusters of primordial black holes.
\end{abstract}

\date{\today}
\maketitle

\section{Introduction}
The QCD axion, a leading solution for the strong CP problem, can also be the dark matter if its mass falls in the range $1-100\mu$eV \cite{PhysRevLett.40.223,PhysRevLett.40.279,PhysRevLett.43.103,Abbott:1982af, Dine:1982ah,Preskill:1982cy, Peccei:2006as}, and possibly smaller masses still with anthropic tuning \citep{2004hep.ph....8167W} or a low inflation scale \citep{2018PhRvD..98c5017G}. In the post-inflationary scenario where Peccei-Quinn symmetry \citep{Peccei:1977hh} is broken after inflation, axion dark matter has enhanced small-scale structure that may make it more observable.  This enhanced structure occurs because the axion field is coherent over the Hubble horizon when Peccei-Quinn symmetry is broken. Afterwards, the axion field is smoothed by its dynamics on scales up to the horizon size, plus additional discontinuities from topological defects \cite{Kibble:1976sj}. The axion then acquires its mass during QCD era and starts to act as nonrelativistic matter. 
The local number density of axions when the axion acquires its mass is directly determined by the axion field value there, leaving primordial inhomegeneity at the horizon scale. 
Topological defects from the symmetry breaking like axion strings also introduce inhomogeneities, extending axion perturbations to even smaller scales   \cite{Vaquero:2018tib,Gorghetto:2020qws}.

After the axion becomes nonrelativstic, gravity takes over as the dominant force. Axions residing in $\gtrsim 1$ fractional density perturbations will collapse at matter-radiation equality and form `axion miniclusters' with characteristic masses of $\sim 10^{-12}M_{\odot}$ and radii of $\sim 10^{12} \rm cm$ \cite{HOGAN1988228,Kolb:1993zz,Kolb:1995bu}. (The mass and radii range can be modified by nonstandard thermal history before big bang nucleosynthesis \cite{Nelson:2018via,Visinelli:2018wza}.) 
In the standard thermal history, the mass range of axion substructures might be detected in femto-, pico- \cite{Kolb:1995bu}, and microlensing surveys if the concentration number is larger than $\sim 10^7$ at the characteristic mass \cite{PhysRevD.97.083502}. Those initial miniclusters merged and formed bigger and less concentrated structures after matter-radiation equality. As the axion density perturbations have a white spectrum that extends to long wavelengths, at progressively later times longer wavelength perturbations collapse and form more massive structures. We call these late-time structures axion minihalos, and the low-redshift spectrum of structures that results is what is relevant for proposed observables.  Many of these still form well before the halos of standard inflationary perturbations, and as such the axion minihalos can be much more compact and denser -- making them more hardy to disruption processes and also more detectable.

There are several proposals for detecting axion minihalos.  One uses the Shapiro time delays and Doppler shifts that these axion minihalos could impart on pulsar timing \cite{Dror:2019twh,Ramani:2020hdo}. Pulsar timing signal of dark matter substructures in mass range $10^{-11}$-$10^{3}M_{\odot}$ \cite{Dror:2019twh,Ramani:2020hdo} can be detected in the future by pulsars detected with the Square Kilometer Array (SKA) \cite{Rosado:2015epa}.  A second proposal uses the effect of these minihalos on the microlensing caustics of cosmological stars that are highly magnified both by a cluster lens and stellar microlens \cite{Dai:2019lud}. Such highly-magnified stars have been discovered recently with the Hubble Space telescope  \cite{Kelly2018,Chen:2019ncy,Kaurov:2019alr}, and the James Webb Space Telescope (JWST) is projected to find more of these extreme events \cite{Dai:2018mxx}. Other proposals include detecting radio emission from axion stars (which are likely to form in the central cusp of axion halos \cite{Kolb:1993hw,Kolb:1993zz, Kirkpatrick:2020fwd,Eggemeier:2019jsu,Levkov:2018kau,Bai:2017feq, Hertzberg:2018zte,Hertzberg:2020dbk, Prabhu:2020pzm,Prabhu:2020yif,Carenza:2019vzg}), transients such as Fast Radio Bursts as a potential signal of the explosive decay of axion miniclusters \cite{Tkachev:2014dpa,Iwazaki:2014wka,vanWaerbeke:2018nyj,Dietrich:2018jov,Sun:2020gem,Buckley:2020fmh}, and axion-photon conversion in neutron stars \cite{Safdi:2018oeu,Battye:2019aco, Foster:2020pgt,Hook:2018iia} or the Galactic Center \cite{Caputo:2018vmy}.   Finally, the clumpiness of cosmic axions could affect direct detection efforts like the Axion Dark Matter eXperiment (ADMX)\cite{PhysRevLett.124.101303}, especially if a significant fraction of present-day axions are bound in minihalos.

To place meaningful constraints on axion minihalos and the post-inflation axion scenario requires an understanding of the mass spectrum of these minihalos.
Several works have used semi-analytic models that were developed for the inflationary fluctuations to compute the mass function and concentration of axion minihalos \cite{Dai:2019lud, Enander:2017ogx,PhysRevD.97.083502,Ellis:2020gtq,Blinov:2019jqc}. It is important to test these semi-analytic models using numerical simulations for the much different case of white isocurvature fluctuations that arise for post-inflationary axions.  In addition, the fraction of axion dark matter that collapsed into axion minihalos can only accurately be estimated by numerical simulations. The spectrum of axion density perturbations prior to the QCD epoch has been numerically simulated \cite{Vaquero:2018tib,Buschmann:2019icd}.  Starting from the predictions of these early universe simulations \cite{Zurek:2006sy, Eggemeier:2019khm}, the subsequent nonrelativistic phase in which the evolution is mainly gravitational has also been simulated, and this work follows this program.  We compute the (subhalo) mass function and density profiles of axion minihalos at all redshifts to $z=0$. 
 Even though our simulation is designed for QCD axions, the initial density power spectrum in our numerical experiments is scale-invariant, which makes generalizing some of our results to other axion-like particles (such as fuzzy dark matter \cite{2010PhRvD..81l3530A, 2017PhRvD..95d3541H}) straightforward.

 This paper is organized as follows: In section \ref{sec:simulation}, we discuss the initial conditions of axion perturbations and other simulation setup. In section \ref{sec:mass_function}, we present the halo mass function obtained from simulation data and fit it with a semi-analytic formula. In section \ref{sec:density_profile}, we present the density profiles and the mass-concentration relation obtained from our simulation data. In section \ref{sec:observations}, we discuss the observations that will be sensitive to those objects. Our simulation is run with a $\Lambda$CDM cosmology with $h$=0.697, $\Omega_{\rm m}=0.2814$, and $\Omega_{\Lambda}=0.7186$, and some of our semi-analytic calculations further adopt $n_s= 0.9667$, which are consistent with Planck cosmic microwave background results \cite{Aghanim:2018eyx}. The radiation component is included in the background evolution with a present-day CMB temperature 2.7255 K and we treat the neutrinos/antineutrinos as massless particles with 3.045 effective degrees of freedom. 

\section{Simulation of Axion Minihalos}\label{sec:simulation}

\begin{figure}
\centering
\includegraphics[width= 0.5\textwidth]{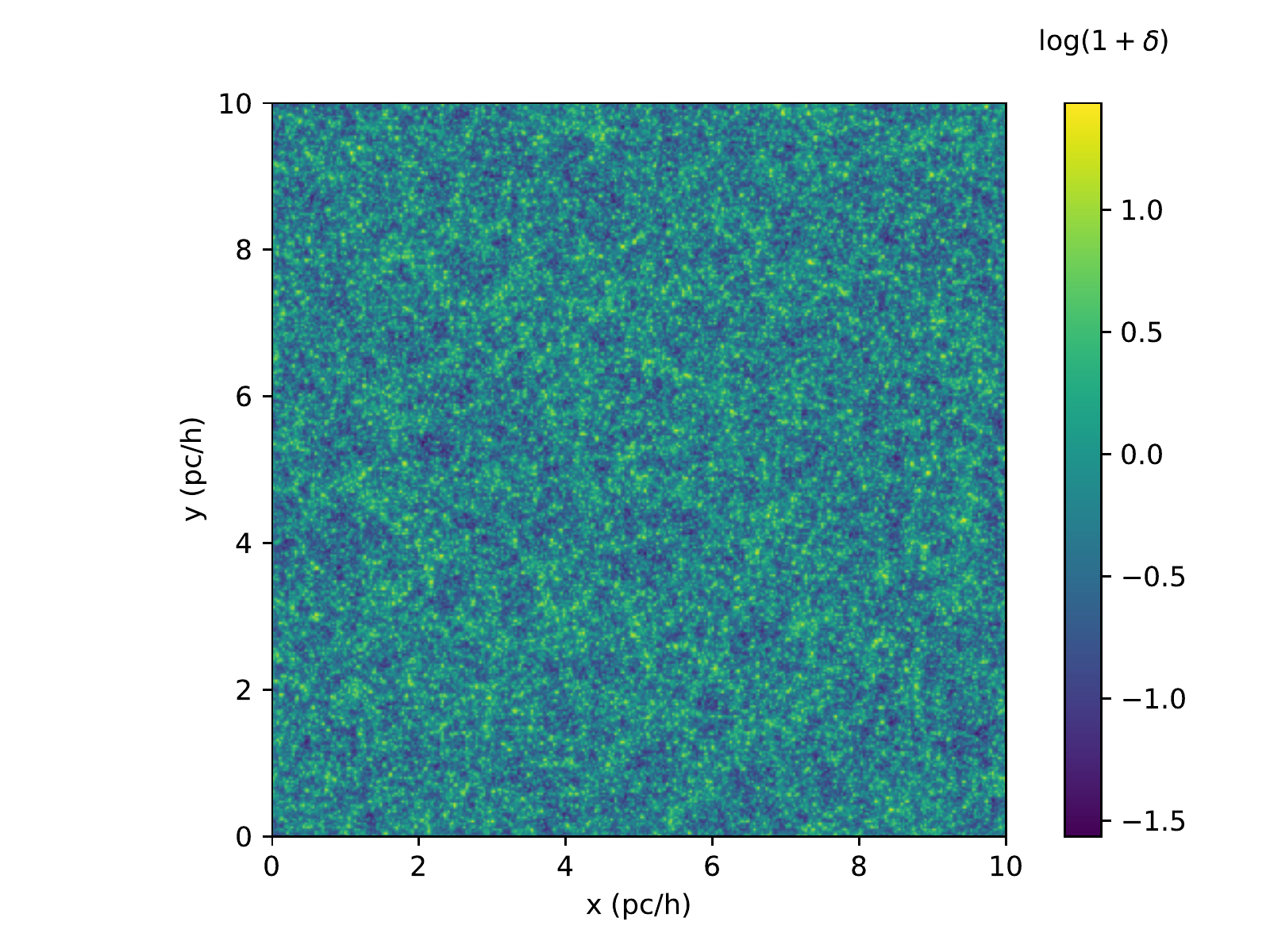}
\includegraphics[width= 0.5\textwidth]{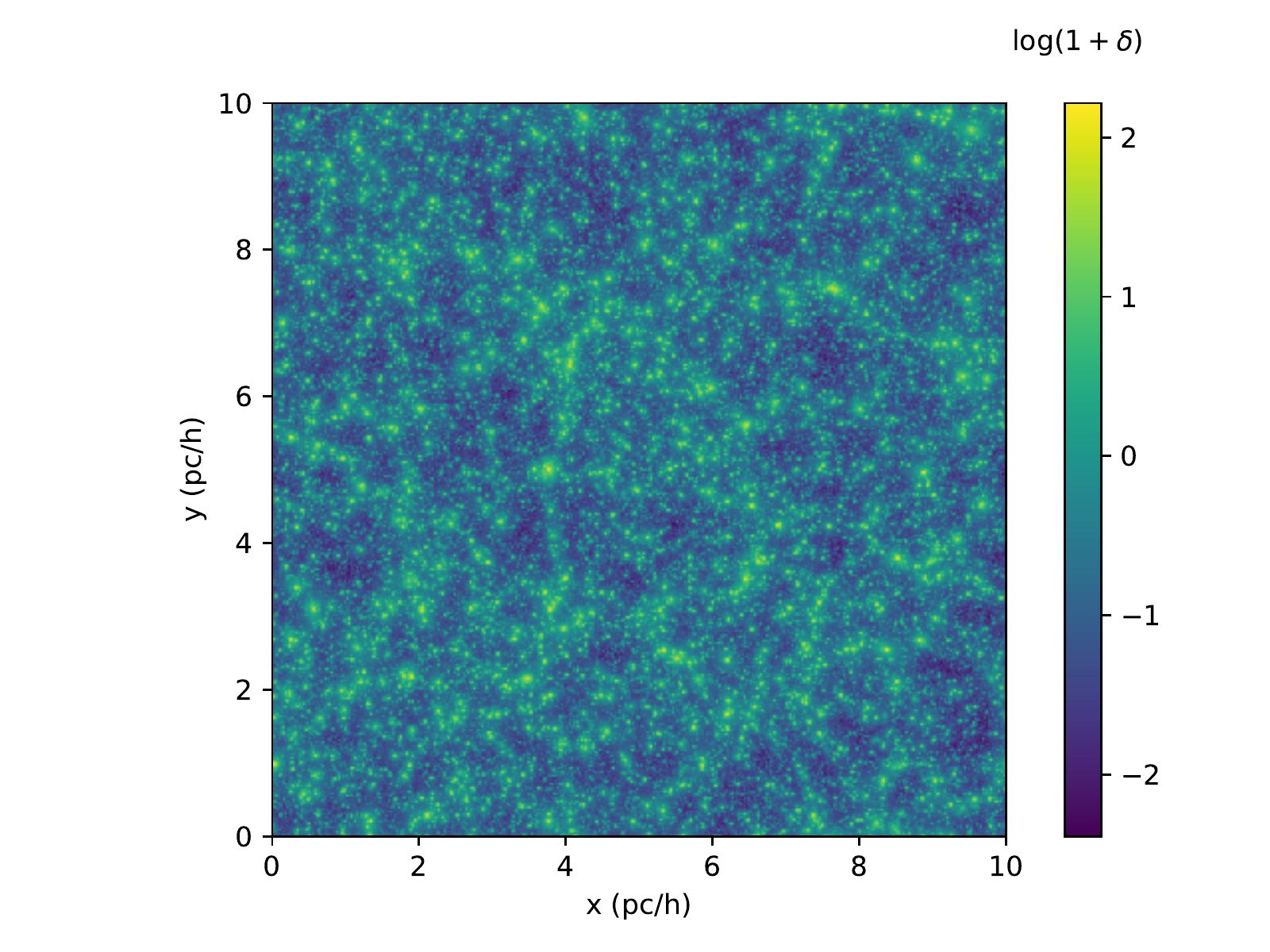}
\caption{Visualization of a projection through the $10/h$ pc simulation at $z=99$ (top panel) and z=24 (bottom panel). The color scale indicates the logarithm of the projected density field, $\log(1+\delta_P)$, where $\delta_P$ is the projected overdensity.
The size of axion minihalos grow significantly between these two redshifts, yet they do not inhabit as distinctive a cosmic web as lower redshift halos that form from the adiabatic perturbations from inflation.}
\label{fig:simulation_vis}
\end{figure}

We follow the nonlinear gravitational evolution of the initial axion perturbations with the MP-Gadget code \cite{yu_feng_2018_1451799}, which is based on GADGET-2 \cite{2005MNRAS.364.1105S}. 
 Our simulations focus on larger axion minihalos than previous simulations (such as \cite{Eggemeier:2019khm}).  This choice is motivated by our finding that most of the mass at low redshifts is in relatively massive $\sim 10^{-7}M_{\odot}$ minihalos.  This mass is much greater than the mass of the first axion minicluster halo that forms, which are smaller than the mass within the horizon when the axion becomes nonrelativistic ($\sim 10^{-12}M_{\odot}$).  This motivates a simulation box size that is much greater than this horizon.  A second motivation is that the axion isocurvature perturbations are white on these super-horizon scales, whereas on smaller scales is set by complicated dynamics that requires evolving the relativistic sine-Gordon equation. 

Our simulations start at $z=30,000$ with radiation background included in the expansion. The evolution of axion isocurvature perturbations is dominated by the matter density and the radiation component can be approximated as a uniform background. 
The comoving size of our fiducial simulation is 50 pc$/h$, and the simulation follows the gravitational dynamics of $1024^3$ dark matter particles each with a mass of $1.3\times 10^{-11}M_{\odot}$.  We also run a smaller 10pc$/h$ box with $512^3$ particles to test a different power spectrum cutoff at high $k$, as well as a series of 50 pc$/h$, $512^3$ simulations to test convergence in force softening and time stepping (Appendix~\ref{appendix:convergence}).  However, as our simulations are largely scale invariant with structure formation occurring in a self-similar manner during matter domination, they can essentially be remapped to other mass scales to study other axion-like scenarios.  All simulations are run to $z=19$. By this point, larger scales than captured in our box start to collapse and so it is not motivated to run the simulation further.  However, we devise  semi-analytic tools to predict the subsequent evolution of minihalos in and outside of more massive halos.

Our results only sensitively depend on the spectrum of density fluctuations on separations larger than the comoving Hubble scale when the axion starts to oscillate in its potential and act as nonrelativistic matter, $ {\cal H}_{\rm osc}^{-1}$, where ${\cal H} = aH$ and $a$ is the scale factor and $H$ is the Hubble function.  
 Resolving the initial perturbations on smaller scales would not change our results significantly, as the collapse on larger scales effectively erases the smaller  collapsed structures that initially form.  However, these smaller structures, if sufficiently segregated in mass and collapse redshift, would likely be subhalos in the larger axion minihalos that later form.

The axion isocurvature fluctuations at wavenumbers with $k \ll {\cal H}_{\rm osc}$ follow a white noise power spectrum because different ${\cal H}_{\rm osc}^{-1}$ patches were not in communication and so there can be no correlation.  Furthermore, because $\gg 1$ horizon patches contribute to these scales, we assume the spectrum of fluctuations is Gaussian so that the power spectrum full describes the fluctuations, although we further justify this approximation later.  We initialize our simulation with a white power spectrum until a sharp cutoff:
\begin{equation}\label{eq:power}
    \Delta^2_{\mathcal{L}}(k)\equiv \frac{k^3}{2\pi^2}P_{\mathcal{L}}(k)= A_{\rm osc} \left(\frac{k}{k_{\rm osc}}\right)^3 \,\,\,\,\, {\rm at}\,\, k<k_{\rm osc},
\end{equation}
where $P_{\mathcal{L}}(k)\equiv V^{-1}|\tilde{\delta}_{\mathcal{L}}(\mathbf{k})|^2$, $V$ is the volume, $\tilde{\delta}_{\mathcal{L}}$ is the Fourier transform of the configuration-space linear dark matter overdensity, $k_{\rm osc} \equiv  {\cal H}_{\rm osc}$ and can be phrased in terms of the axion mass at low temperatures, $m_a$, via \cite{Vaquero:2018tib}
\begin{equation}\label{eq:axion_mass}
    \frac{1}{k_{\rm osc}}=0.036\left(\frac{50{\rm \mu eV}}{m_a}\right)^{0.17} {\rm pc},
\end{equation}
 and $A_{\rm osc}$ sets the normalization, with $A_{\rm osc}\sim 1$ resulting in order-unity fluctuations on the oscillation scale \footnote{ In contrast to the small scale limit of $\rm \Lambda$CDM model at late times, the dimensionless initial (post QCD-epoch) linear dark matter power spectrum $\Delta_{\mathcal{L}}^2$ in our axion cosmology can reach ${\cal O}(1)$ at high wavenumbers, indicating that small scales will collapse as early as matter-radiation equality.  However, for the mass scales we consider, the collapse occurs generally well after matter-radiation equality}. Simulations of the QCD axion find that values of the
isocurvature variance at initial conditions are $A_{\rm osc}\sim 0.01 -0.3$ \cite{Vaquero:2018tib, Buschmann:2019icd}, where the large range of values we think owes to the importance of axion strings over the misalignment mechanism in establishing the density fluctuations. (There is some controversy in the importance of radiation from axion strings, as discussed later.  Our results are applicable regardless.)  Though we start our simulation well after the time when the field has become nonrelativistic (at $z=30,000$), Eq.~\ref{eq:power} still holds to good approximation at subsequent times during radiation domination \citep{Irsic:2019iff}.  

Eq.~\ref{eq:power} has an artificially sharp cutoff in the spectrum at $k_{\rm osc}$.  In the misalignment mechanism for density perturbations, a somewhat sharp cut-off is expected for the modes that entered the horizon when the axion behaved relativistically, as the field homogenize at higher wavenumbers.  Defects such as axion strings can result in an even weaker cutoff.  However, for the mass scales probed by pulsar timing and microlensing lensing observables that we focus on, the precise wavenumbers and spectral shape of the cut-off are irrelevant since our results will not be sensitive to the affected scales. Therefore, our power spectrum is effectively scale-invariant for the dynamical range of interest. One may also worry about the axion Jeans scale, $k_{\rm J}$, where density and pressure are in equilibrium. Modes can grow only when $k<k_{\rm J}$ otherwise it will be smoothed by pressure. The Jeans scale is given by $k_{\rm J}\approx 2\times 10^4a^{1/4}\sqrt{m/(10^{-5}\rm eV)} \rm pc^{-1}$ if the axion is the dark matter \cite{Marsh:2015xka}. Therefore, the modes resolved in our simulation box satisfy $k<k_{\rm J}$. 


The initial conditions of our simulations are generated with the parameters $A_{\rm osc}=0.1$ and the momentum cutoff to be $k_{\rm osc}=19.8~ \rm pc^{-1}$, corresponding to axion mass $m_a=6.9\mu\rm eV$ and decay constant $f_a\approx 10^{12}\rm GeV$. Note that the axion decay constant plays no role in our simulation. The only parameter that matters for structure formation on most mass scales is the amplitude of the power spectrum, $A_{\rm osc}/{k_{\rm osc}}^3$.  There are some uncertainties on axion mass due to the production of axions from axion string decay, leading to different $m_a$ and $k_{\rm osc}$ \cite{Gorghetto:2020qws}. We use our simulations to calibrate a semi-analytic model in \S~\ref{fig:mass_function}, which allows us to model a broad class of post-inflation axion scenarios.

All modes in our simulation are well within the Horizon at the starting time of the simulation and have a growing and decaying component. We only include the growing mode because of the following justification: The comoving scale of perturbations that contribute to the halos we focus on is larger than $1/k_{\rm osc}$.  On such scales the isocurvature fluctuations are much smaller than one ($\Delta^2_{\mathcal{L}}(k)\ll1$), meaning that the formation of axion structures occurs after matter-radiation equality when the decaying mode is being redshifted away.   The picture for isocurvature fluctuations contrasts with adiabatic fluctuations, where the overdensity contrast grows outside the Horizon, at least in the Newtonian gauge, and then its growth freezes if it enters the horizon during radiation domination, leading to additional scale dependence depending on when a mode enters the horizon. This growth outside the horizon is not present for our isocurvature fluctuations.  Once our isocurvature perturbation is within the horizon so that the perturbations in radiation and baryons have been damped, the growing mode of the dark matter in radiation- and matter-dominated epochs can be well described by a solution of the M$\rm\Acute{e}$sz$\rm\Acute{a}$ros equations:
\begin{equation}\label{eq:growth_function}
    D = \frac{2}{3}+\frac{a}{a_{\rm eq}},
\end{equation}
where $D$ is the growth function of cosmological perturbations. At late times when the dark energy starts to dominate, there are corrections to this growth function that are included in this study.  In \cite{Irsic:2019iff}, the scale-independent linear evolution given by Eq.~\ref{eq:growth_function} was checked against a full Boltzmann calculation with isocurvature perturbations and found to agree well. (See this footnote for additional discussion: \footnote{An approximation is that our simulations are in the standard limit where all the matter is dark matter, there is a factor of $\Omega_{\rm DM}/\Omega_m \approx 0.8$ suppression over the growth rates relatively to our simulation.  The large uncertainty in $A_{\rm osc}$ justifies omitting this factor.}) We tested the power spectrum in the simulation at early times and found that its behavior is well described by this growth factor in the linear regime.


The initial condition of our simulation is generated with the MP-GenIC code included in MP-Gadget which uses linear order Lagrangian Perturbation Theory (1LPT). Displacements and velocities are generated assuming the power spectrum given by Eq.~\ref{eq:power} with all the power growing as Eq.~\ref{eq:growth_function}.  We justify this perturbative approach because $\Delta^2_{\mathcal{L}}(k)\ll1$, reaching a maximum of $\Delta^2_{\mathcal{L}}(k_{\rm osc}) = 0.1$. Since the initial time we start the simulation is way after the QCD era, axions are already nonrelativistic at the scale of interest and the velocity is only slightly nonzero in the 1LPT treatment. The initial conditions in Ref. \cite{Eggemeier:2019khm} are generated with zero initial velocity. 

The simulation then evolves the linear axion perturbations from $z=30,000$ to where they become nonlinear and form of axion minihalos. Substructure forms at much earlier times than in the standard $\Lambda$CDM cosmology. A visualization of our simulation at $z=24$ and $z=99$ is shown in Fig.~\ref{fig:simulation_vis}.

\section{Mass Function of Axion Minihalos}\label{sec:mass_function}

\begin{figure*}
\centering
\includegraphics[width= 0.9\textwidth]{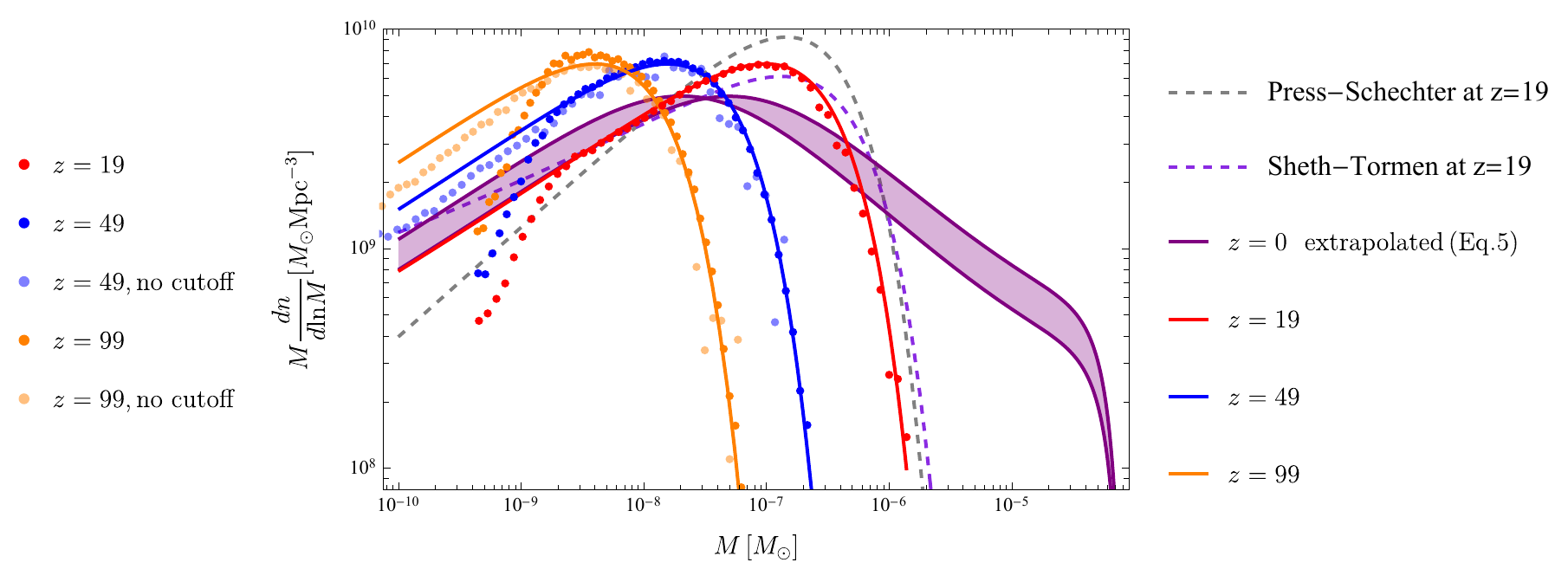}
\caption{The mass function of axion minihalos at select redshifts for a white power spectrum plus cutoff specified by $A_{\rm osc}=0.1$ and $k_{\rm osc}=19.8~ \rm pc^{-1}$. We compare mass functions obtained from two set of simulations to those in semi-analytic models. The data points are from simulations, and the solid curves are the semi-analytic formula discussed in Appendix~\ref{appendix:Fitting}, which recalibrates the barrier parameters of the Sheth Tormen mass function.  We also plotted the Press-Schechter and Sheth-Tormen mass functions at $z=19$. The recalibrated model agrees with the original simulation with a cutoff in the power spectrum except for the high mass end. Another simulation that does not have a cutoff in the power spectrum, denoted by ``no cutoff'', agrees with more broadly with the simulation. 
 The solid ``z=0 extrapolated (Eq.~\ref{eq:current_mf})'' band is our model for the present-day number density of axion minihalos.  It is our estimate for the mass function of minihalos that resides in most dark matter halos, including those that exist as subhalos.
We assume axion minihalos will fall into larger CDM halos (that are too large to be captured in our simulation) and stop merging. The width of the purple band is set by whether the growth of axion minihalos is terminated by when the larger host halo collapses or when it is at turnaround.}
\label{fig:mass_function}
\end{figure*}

 We use the Friends-of-Friends (FOF) algorithm \cite{1985ApJ...292..371D} to identify groups of particles in the simulation, demanding any particle that finds another particle within a linking-length distance of $l$ is linked to it to form a group. We choose $l = 0.2\, d$, where $d$ is the mean separation of dark matter particles in the simulation. The minimum number of dark matter particles in each FOF groups is chosen to be $N_g = 32$.  This $N_g$ corresponds to a minimum halo mass of $4\times10^{-10} M_\odot$ in our fiducial simulation. These choices of $l$ and $N_g$ are standard in cosmological studies, as they approximately select groups that are large enough to be reliably captured in the simulation and link regions with overdensity greater than $80$ \cite{2011MNRAS.415.2293K}, characteristic of the outskirts of dark matter halos. (Minihalos that form during matter domination have a characteristic overdensity of $\sim 200$ at the time of formation.)

 Studies of dark matter halos in the standard CDM cosmology have found great success at explaining the mass function of halos with semi-analytic models developed in the excursion set formalism \citep{1991ApJ...379..440B}.  It is unclear how well these models should work in our case, where the density perturbations have a much bluer spectrum, but success for other nonstandard cosmologies gives reason to believe these models may work for our blue spectrum as well \citep{2009arXiv0908.2702B}. We use these semi-analytic formulas to fit the mass function obtained from the simulation.  In particular, we consider Press-Schechter \cite{1974ApJ...187..425P}, Sheth-Tormen \citep{1999MNRAS.308..119S}, and a tweaked mass function that changes the barrier parameters in Sheth-Tormen.  See Appendix~\ref{appendix:Fitting} for the relevant formulae.  We note that in the standard CDM cosmology, Press-Schechter is able to explain the halo mass function at the factor of $\sim 2$ level (crudely speaking, as the differences are exponentially enhanced at the highest masses), whereas Sheth-Tormen has been tuned to give percent-level agreement. For the post-inflation axion scenario, both disagree with the simulation at the factor of $\sim 2$ level, whereas our tweaked Sheth-Tormen mass function fits well the simulated mass function, except for the masses influenced by our wavenumber cutoff at $k_{\rm osc}$ ($M\lesssim 10^9 M_{\odot}$). 
 
The halo mass function of axion minihalos computed from our fiducial simulation is shown in Fig.~\ref{fig:mass_function}.  The y-axis shows the comoving number density of axion minihalos  per logarithmic mass interval and, then, multiplied by halo mass, such that the integral over this quantity in $\ln M$ gives the total mass in minihalos. As time evolves, the mass function shifts towards higher masses as halos grow and merge. At all redshifts there is a peak halo mass around which much of the mass in axion minihalos lies (e.g. $M\sim 10^{-7}M_{\odot}$ at $z=19$), although there is a substantial fraction of lower mass halos.
 
 The shape of the mass function at $M\lesssim 10^{-9} M_{\odot}$ is affected by the cutoff in our (input) linear power spectrum at $k_{\rm osc}$.  A wavenumber of $k_{\rm osc}$ can be turned into a characteristic mass scale \cite{Dai:2019lud}:
\begin{equation}\label{eq:M0}
  M_0 \equiv \frac{4\pi}{3}\left(\frac{\pi}{k_{\rm osc}}\right)^3  \bar{\rho}_{a,0}=2.3\times 10^{-10}\left(\frac{50\mu\rm eV}{m_a}\right)^{0.51}M_{\odot},
\end{equation}
where $\bar{\rho}_{a,0}$ is the mean density of axions in the present day. For our initial conditions, $M_0 = 6.3\times 10^{-10}M_{\odot}$, corresponding to axion mass $m_a=6.9\mu\rm eV$.  We have run simulations without a cutoff at $k_{\rm osc}$ (`no cutoff' in Fig.~\ref{fig:mass_function}, with 10pc and $512^3$ particles).  Comparison of the simulations with and without a cutoff shows that the break in the small-scale power-law scaling of the mass function manifests at several times larger scales than $M_0$.  Interestingly, the semi-analytic model only shows a break around $M_0$ (a feature that falls off the left-hand side of the plot), so the effect of a cutoff in the power spectrum on the mass function is not well captured in these excursion set models for the mass function.
However, as the bulk of the mass in dark matter halos at $z<100$ resides in halos more massive than are affected by this break, the mass scales around where the break manifests are not important for our primary results.  Indeed, a scale-invariant power spectrum is a sufficient description at almost all scales we study: the only relevant parameter is the power spectrum amplitude $A_{\rm osc}/k_{\rm osc}^3$.



\begin{figure}
\includegraphics[width= 0.5\textwidth]{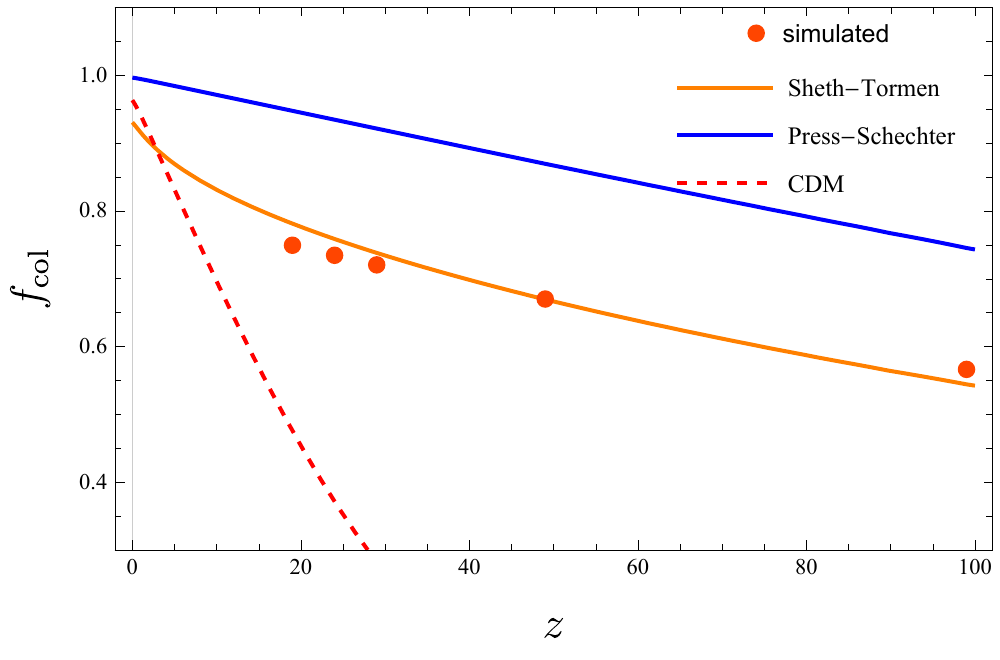}
\caption{The collapsed fraction of axion dark matter as a function of redshift for $A_{\rm osc}=0.1$ and $k_{\rm osc}=19.8~ \rm pc^{-1}$. The orange data points are the collapsed fraction computed as the mass fraction in our FOF groups from our fiducial white-noise simulation, while the orange and blue curves are the collapse fraction of white noise perturbations in the Sheth-Tormen model and Press-Schechter model, respectively. The collapsed fraction in the Sheth-Tormen and the Press-Shechter models is obtained by intergrating over our white noise mass functions from  $M_0= 7\times 10^{-10}M_{\odot}$ to $10^{-4}M_{\odot}$.  The red dashed curve is the Press-Schechter collapse fraction in `CDM halos' that form from adiabiatic fluctuations (which are calculated by cutting off to include only masses of $M>10^{-3}M_\odot$ halos for which adiabatic fluctuations dominate).  This collapse fraction which appears in our froward-evolution model (Eqs.~\ref{eq:current_mf} and \ref{eqn:fcol}).
}
\label{fig:collapsed_fraction}
\end{figure}

Finally, we calculate the total amount of bound structure in axion minihalos. This fraction is important for direct detection efforts, which are sensitive to the unbound component. The fraction of axion dark matter in these bound structures is found to be $0.8$ at z=19, growing from $0.6$ at $z=100$.  This is smaller than in Press-Schechter but comparable to the Sheth-Tormen mass function (see Fig.~\ref{fig:collapsed_fraction}). We expect that this estimate for the total mass in minihalos is an upper bound, especially in a Milky Way-like environment where stellar disruption processes are important (see \S~\ref{sec:disruption}).

\subsection{Comparison with mass function calculated from early universe axion simulations}

\begin{figure}
\centering
\includegraphics[width=0.5\textwidth]{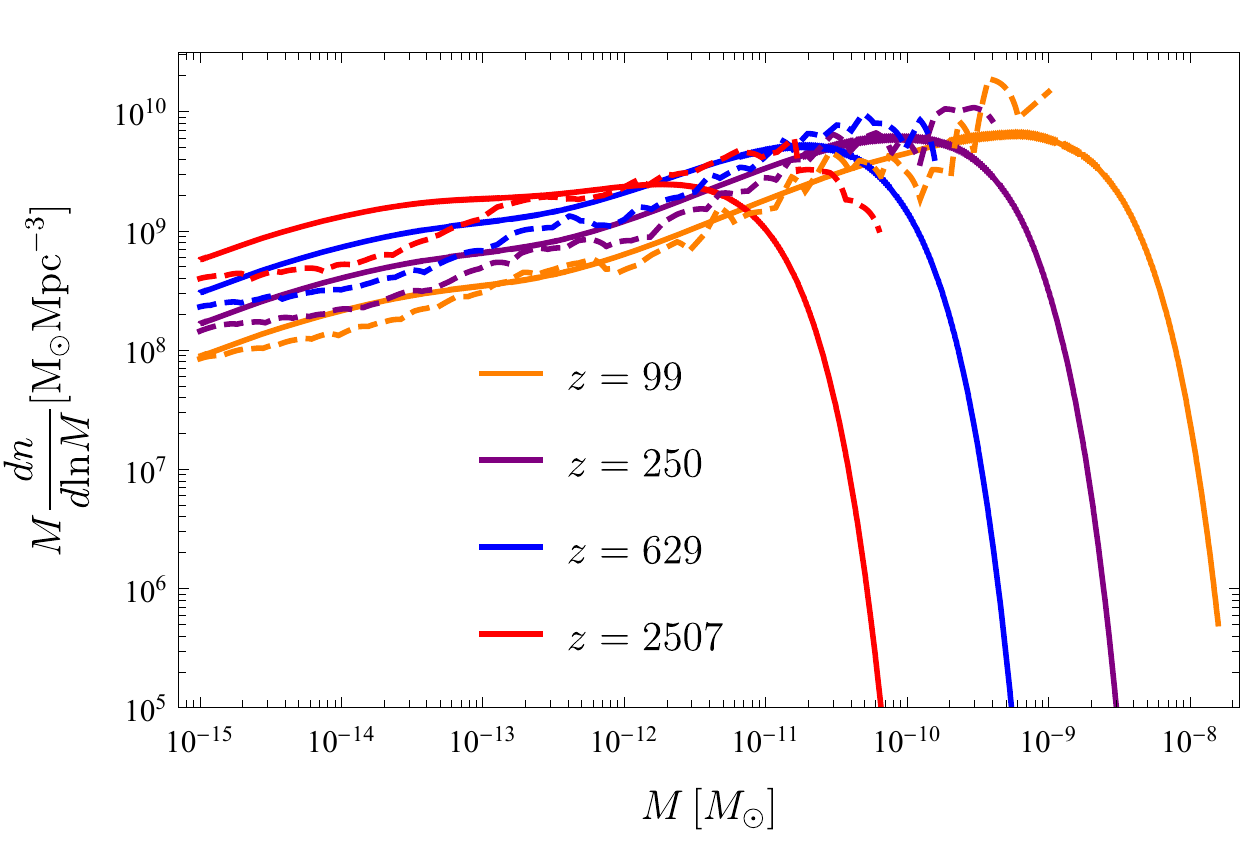}
\caption{Comparison of the mass function of axion minihalos from \cite{Eggemeier:2019khm} with the prediction of our semi-analytic halo model using the same initial density power spectrum.  The semi-analytic model is tuned to fit our simpler white-noise simulations, but still describes reasonably the simulated mass function of \cite{Eggemeier:2019khm}.  The N-body simulations in \cite{Eggemeier:2019khm} evolve the density field predicted from detailed early universe  simulations of the relativistic axion field, which results in the initial conditions for their N-body simulations having a less abrupt cutoff in the power spectrum at high wavenumbers than in our simulations. However, our analytic model broadly agrees, especially at lower redshifts (we suspect because the non-gaussianty in the field matters less at these times).  This agreement allows us to forward model their predictions, which yields at late times curves similar to our white noise model shown in  Fig.~\ref{fig:current_mf}.
\label{fig:mass_functon_comparison}}
\end{figure}

Previous N-body simulations have attempted to start from the initial conditions calculated from solving the sine-Gordon equation for the axion \citep{Zurek:2006sy, Eggemeier:2019khm}.  Unlike our simulations, which are of the white spectrum that occurs at $k\ll {\cal H}_{\rm osc}$, their concentration was on $k\gtrsim {\cal H}_{\rm osc}$, scales shaped by the early universe axion dynamics. Such a study was recently reported in \cite{Eggemeier:2019khm}, whose halo mass function at different masses is shown by the dashed lines in Fig.~\ref{fig:mass_functon_comparison}.  We use their results here to understand whether the tools we develop are applicable for understanding axion structure formation on the scales they studied. 

In particular, we have evaluated our semi-analytic model, calculated using the input power spectrum for the simulations (solid curves in Fig.~ \ref{fig:mass_functon_comparison}).  At late times, this model is generally able to reproduce the evolution seen in their N-body simulation.  The model errs most for the highest redshift \cite{Eggemeier:2019khm} reported, $z=2507$. Our semi-analytic model is motivated by matter-dominated spherical collapse calculations that do not apply during radiation domination and need some modification. For example, in Press-Schechter, the barrier $\delta_{\rm c}$ is no longer 1.686 once radiation is important. While our model should not be used when radiation is important in the universe, improving our model to extend into radiation domination would further enhance the differences.  Rather, the mass function found in \cite{Eggemeier:2019khm} at these times is likely driven by the collapse of small-scale non-Gaussian structures that our Gaussian model does not capture (the formalism from which these semi-analytic models are built is based on Gaussian random walks).  We surmise that the agreement of the simulations of \cite{Eggemeier:2019khm} and our model at later times occurs for two reasons.  First, for smaller mass scales than the our peak-scale, the model is less sensitive to deviations from Gaussianity since the minihalos that are collapsing are not the rare peaks for which non-Gaussianity has its largest effect.  Further, as time progresses, the mass scale at which halos are rare (exponentially suppressed) moves to larger and larger masses.  Many small-scale contributions sum to determine the perturbations on these larger scales, making the fluctuations on these scales more Gaussian (although note that the model is not well tested by the simulations of \cite{Eggemeier:2019khm} at these high masses).

Finally, we remark that even though our analytic model is able to capture the mass function of \citep{Eggemeier:2019khm} at early times, at late times the amplitude of the white noise power at $k\ll {\cal H}_{\rm osc}$ shapes the mass function -- this study presents precise simulations in this regime. We use the normalization of \citep{Eggemeier:2019khm} at these scales as our fiducial model when we analyze observations in \S~\ref{sec:observations}. The complex spectrum of fluctuations on $k\gtrsim {\cal H}_{\rm osc}$ that were simulated by \citep{Eggemeier:2019khm} should largely affect the subhalo distribution in the larger axion minihalos.

\subsection{Extrapolating the mass function to the present day}
Our simulation boxes, with our largest containing in total a cosmologically minuscule mass of $10^{-2} M_\odot$, are far from capturing the scales needed to follow cosmological structure formation, which results in most of the dark matter being incorporated galactic mass halos of $\gtrsim 10^{10} M_\odot$ today.  Thus, we need a way of incorporating the effects of these large scales.  Fortunately, the separation of mass scales between what later collapses and our axion minihalos suggests a reasonable approach.

An axion halo will stop growing once it falls into CDM halo.  We distinguish ``CDM halos'' as the halos that form from the standard adiabatic perturbations from inflation.  The spectrum of adiabatic density perturbations is such that a large range of masses collapse at around the same time: few CDM halos have formed at $z\sim 50$, but most have collapsed by $z\sim 10$.  Thus, most of the axion minihalos are incorporated into the larger CDM halos at these redshifts.  However, because the host CDM halos are generally much larger, most axion minihalos do not spiral inward from dynamical friction and merge into a single halo.  Furthermore, once subsumed in the larger halo, they stop growing via accretion because of the high velocities in the host halo.  

Therefore, we estimate the mass function at a given redshift of axion minihalos, \emph{including those incorporated into massive CDM halos}, as 
\begin{eqnarray}
\frac{dn_{f}}{dM}(z) &=& \int_{z_{\rm eq}}^{z} dz' \frac{df_{\rm col}^{\rm CDM}(z')}{dz'}\frac{dn_{\rm WN}}{dM}(z') \nonumber \\
&&+  \left[1-f_{\rm col}^{\rm CDM}(z) \right]\frac{dn_{\rm WN}}{dM}(z), 
\label{eq:current_mf}
\end{eqnarray}
where $dn_{\rm WN}/{dM}$ is the mass function computed from our white noise--only simulations and $f_{\rm col}^{\rm CDM}(z)$ is the collapse fraction of CDM halos at redshift $z$ into halos (shown with the dashed curve in Fig.~\ref{fig:collapsed_fraction}), which goes from near zero to near unity over $z\sim 10-50$ -- freezing the spectrum of axion minihalos.  The second term on the right hand side, which comes from axion minihalos unbound to any CDM halos, becomes small at low redshifts. While this estimate provides the global mass function of axion minihalos, including those that are subhalos in CDM halos, Eq.~\ref{eq:current_mf} likely approximates the mass function of axion minihalos at $z=0$ within any Milky Way-like or cluster-scale CDM halo because the $f_{\rm col}^{\rm CDM}$ history of the matter within these much larger halos is not be significantly different than the mean history. This estimate is a different approach from previous studies where the mass function is calculated using the power spectrum from the sum of isocurvature and adiabatic fluctuations \cite{Dai:2019lud,Lee:2020wfn}. In our estimate, we tracked all the minihalos arising from isocurvature fluctuations even though they may have fallen into CDM halos formed from adiabatic fluctuations. We contrast the two methods in the conclusions.

The estimate given by Eq.~\ref{eq:current_mf} may overestimate the fraction of axion minihalos that exist today because it does not account for destruction processes.  Destruction owing to encounters with stars and other minihalos are discussed in \S~\ref{sec:disruption}.  We argue that in galactic environments stellar destruction can remove an {\cal O}(1) fraction, but it's likely many still remain.  Another process is the tidal stripping of the host macro-halo itself on the minihalo.  See this footnote for discussion  \footnote{The tidal radius of a minihalo sitting $\sim 1$ scale radius from the center of the host macro-halo is roughly the scale radius of the minihalo times the ratio of the concentrations of the macro halo and minihalo.  (Tidal stripping is further enhanced by the radial biased nature of orbits in CDM halos.)  Thus more concentrated minihalos are more likely to survive in their host halo.  Because axion minihalos form much earlier than CDM halos, making them more concentrated, one might think we are safe to ignore tidal stripping.  However, our most massive minihalos are our least concentrated, forming at a time similar to CDM halos.  Thus, we expect some tidal stripping will be important, particularly for the most massive minihalos.}.  While beyond the scope of this study, techniques that have been developed for the standard cosmology to evaluate the tidal stripping and survival of subhalos should be applied to evolve axion minihalos \citep{2001ApJ...559..716T,  2020arXiv201107077E}.  

With these caveats in mind, to evaluate Eq.~\ref{eq:current_mf}, we use the Press-Schechter model for the collapse fraction of CDM halos, which yields
\begin{equation}
    f_{\rm col}^{\rm CDM}(z)={\rm erfc}\left(\frac{\delta_c}{\sqrt{2}\sigma_{\rm CDM}(M_{\rm min})D(z)}\right),
    \label{eqn:fcol}
\end{equation}
where $\sigma_{\rm CDM}(M)^2$ is the variance in the initial density fluctuation field, and $M_{\rm min}$ is the smallest halo we are counting in the tally of collapsed structures. We take $M_{\rm min}$ to be $10^{-2}M_{\odot}$, corresponding to a scale where CDM power spectrum starts to dominate and axion perturbations becomes subdominant.  Our results are insensitive to this choice owing to the roughly logarithmic dependence of $\sigma_{\rm CDM}(M)$ on $M$ at relevant masses.  While the Press-Schechter mass function is known to error, the simplicity of our model does not motivate more sophisticated prescriptions for $f_{\rm col}^{\rm CDM}(z)$. We evaluate $\delta_c$ at two values:  Its formal spherical collapse value of $1.69$ and the value for turnaround of $1.06$ (both of these are matter-dominated values, but including dark energy has a percent-level effect by $z=0$).  The two choices quantify some uncertainty in our estimate, as it is unclear whether turnaround or collapse better encapsulates the time when subhalos stop growing owing to tidal effects.  Furthermore, choosing turnaround segregates the time of formation of the host CDM halo from the axion minihalos and, thus, chooses the denser minihalos that form by the earlier turnaround epoch that are more likely to survive.  However, these choices for $\delta_c$ yield results that are within a factor of two of each other.

The solid purple band in Fig.~\ref{fig:mass_function} shows our estimate using Eq.~\ref{eq:current_mf} for the final mass function in axion minihalos.  The upper edge of the purple band at high mass uses $\delta_c=1.69$ and lower $\delta_c=1.06$.  The peak occurs at the peak location of the individual snapshots when much of the mass is incorporated in the standard CDM halos and stop growing (at $z\sim 30$).  However, the mass function extends over a broader range of masses than the mass functions from individual snapshots in our simulations (compare with this curve with, e.g., the $z=19$ curve).  We use this estimate in \S~\ref{sec:observables} to compare with observations.  Since we have an analytic model, we can further take into account uncertainties in the initial spectrum of axion fluctuations.

\section{Density profile of axion minihalos}\label{sec:density_profile}

\begin{figure*}
\includegraphics[width= 1\textwidth]{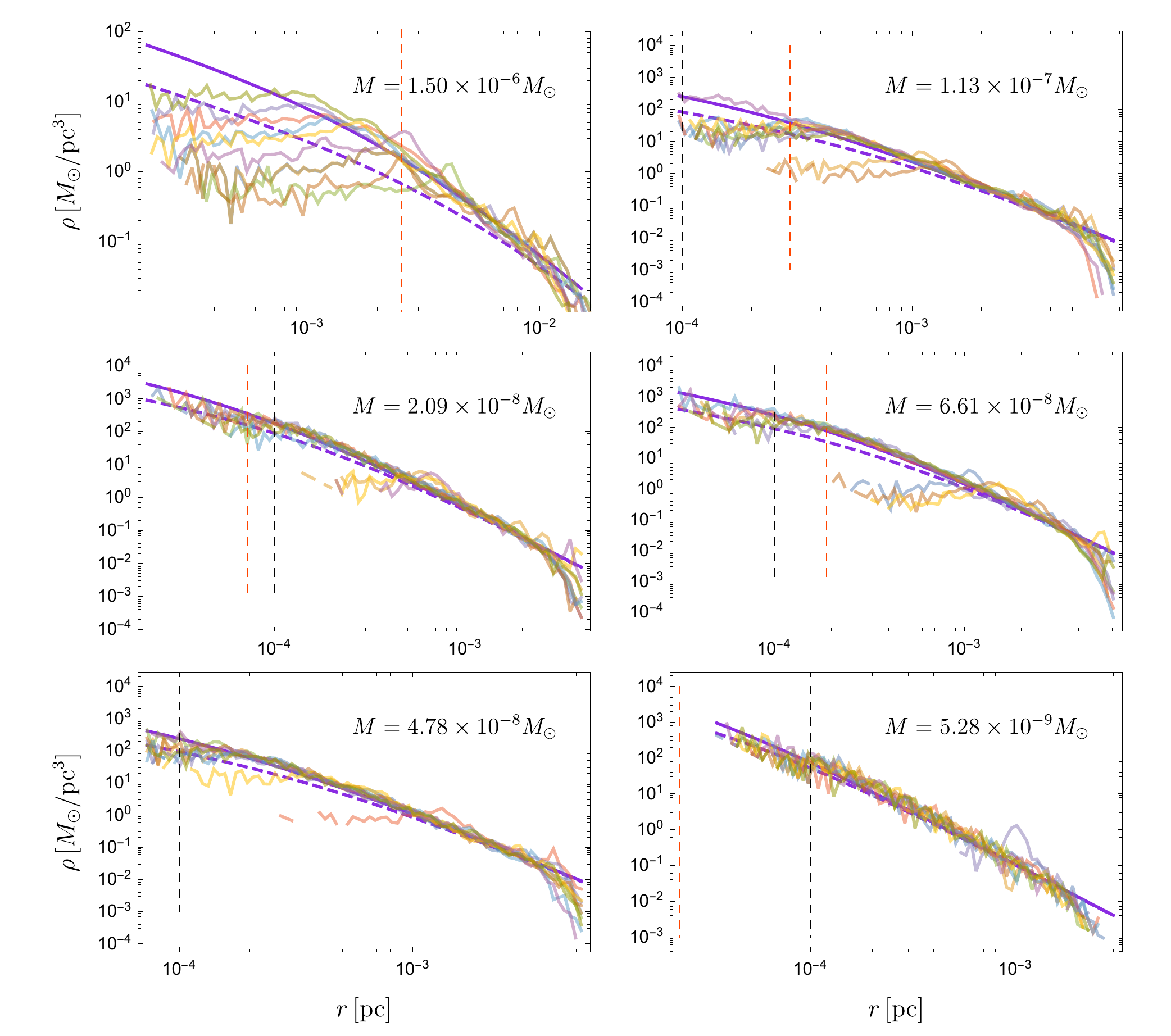}
\caption{Density profile of halos with different masses at $z=19$ from our fiducial simulation interpreted with $A_{\rm osc}=0.1$ and $k_{\rm osc}=19.8~ \rm pc^{-1}$. The colored curves with less opacity represent the density profile of 10 halos around the same mass obtained from our simulation data. The purple solid curve represents the NFW density profile with a mass-dependent scale radius given by Eq.\ref{eq:scale_radius}, whereas the dashed purple curve is re-scaled to a factor of two smaller values for the scale radius (Eq.~\ref{eq:scale_radius}), which suggests how well the scale radius can be determined from the simulation. The vertical black and orange-red dashed line are indicating the gravitational softening length and scale radius of the halo respectively.
The density profiles from simulations broadly agree with the predicted NFW profiles at different masses, except for a contingent of halos that are unrelaxed at their center.  This contingent is especially present the higher masses, and owes to incomplete merging of two halos (Fig.~\ref{fig:halo_vis} shows some examples). 
An NFW profile with the model for the scale radius in Eq.~\ref{eq:scale_radius} agrees reasonably with the more relaxed profiles. 
 }
\label{fig:density_profile_z=19}
\end{figure*}

\begin{figure*}
\includegraphics[width= 1\textwidth]{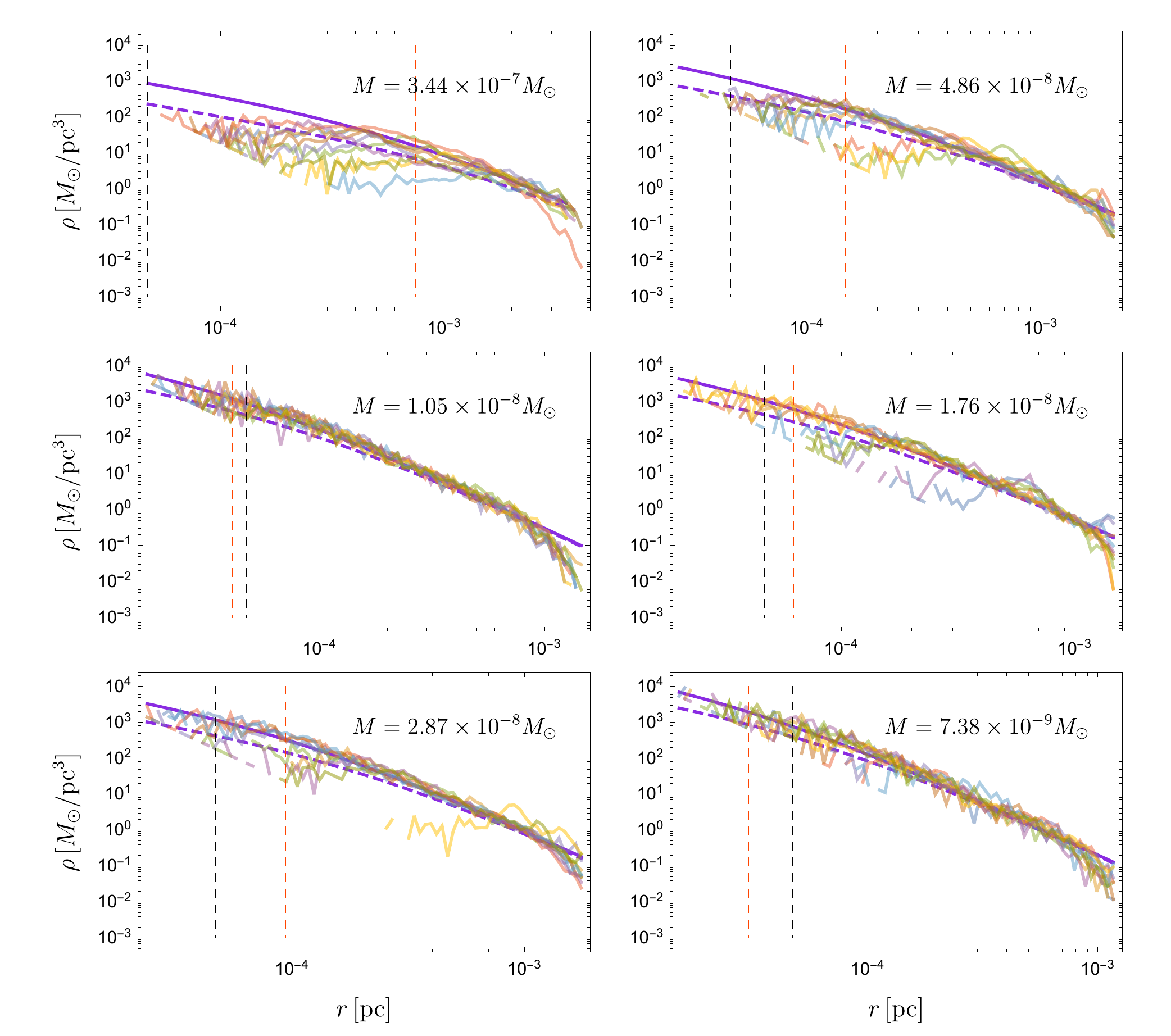}
\caption{The same as Figure~\ref{fig:density_profile_z=19} except showing the density profile of halos with different masses at $z=49$ rather than $z=19$. The colored curves with less opacity represent the density profile of 10 halos around the same mass obtained from our simulation data. The purple solid curve represents the NFW density profile with a scale radius described in Eq.~\ref{eq:scale_radius}, and the purple dashed curve has a scale radius that is a factor of two smaller than the solid curve. The vertical black and orange-red dashed line are indicating the gravitational softening length and scale radius of the halo respectively. The NFW model predictions broadly agrees with the simulation data except for the high mass end where halos have substructures. The main conclusion is the same as what we found in Fig.~\ref{fig:density_profile_z=19}.  In combination the two figures support that the density profile of axion halos do not change over time (at trait that holds also for the NFW model curves). 
}
\label{fig:density_profile_z=49}
\end{figure*}

The density profile of axion minihalos is important for their detectability.  The more dense and concentrated these halos, the more detectable they generally are and the longer they are able to survive in galactic environments.  We study the angular-averaged density profiles at $z=19$ and $z=49$.
Much of our discussion uses the famous Navarro-Frenk-White (NFW) profile and so we start off by defining it and related quantities.  In particular the NFW profile is given by \cite{Navarro:1995iw}:
\begin{equation}
    \rho(r)=\frac{\rho_s}{r/r_s(1+r/r_s)^2},
\end{equation}
where $\rho_s$ is the characteristic density of the halo and $r_s$ is the scale radius.  The scale radius determines the concentration number via $c \equiv r_{\rm vir}/r_s$, where $r_{\rm vir}$ is the virial radius of dark matter halo. The virial radius is defined here by the matter density $\bar{\rho}(z)$ and halo mass $M$ as $4\pi(200) r_{\rm vir}^3\bar{\rho}/3=M$, i.e. the region that encloses an overdensity of $200$. This standard choice is motivated by spherical collapse, and it means that the concentration number, $c$, of a static halo (i.e. one with a fixed density profile) increases with time as $r_{\rm vir} \propto (1+z)^{-1}$.

\begin{figure*}[!tbp]
\centering
\begin{minipage}[b]{0.49\textwidth}
\includegraphics[width= 1.02\textwidth]{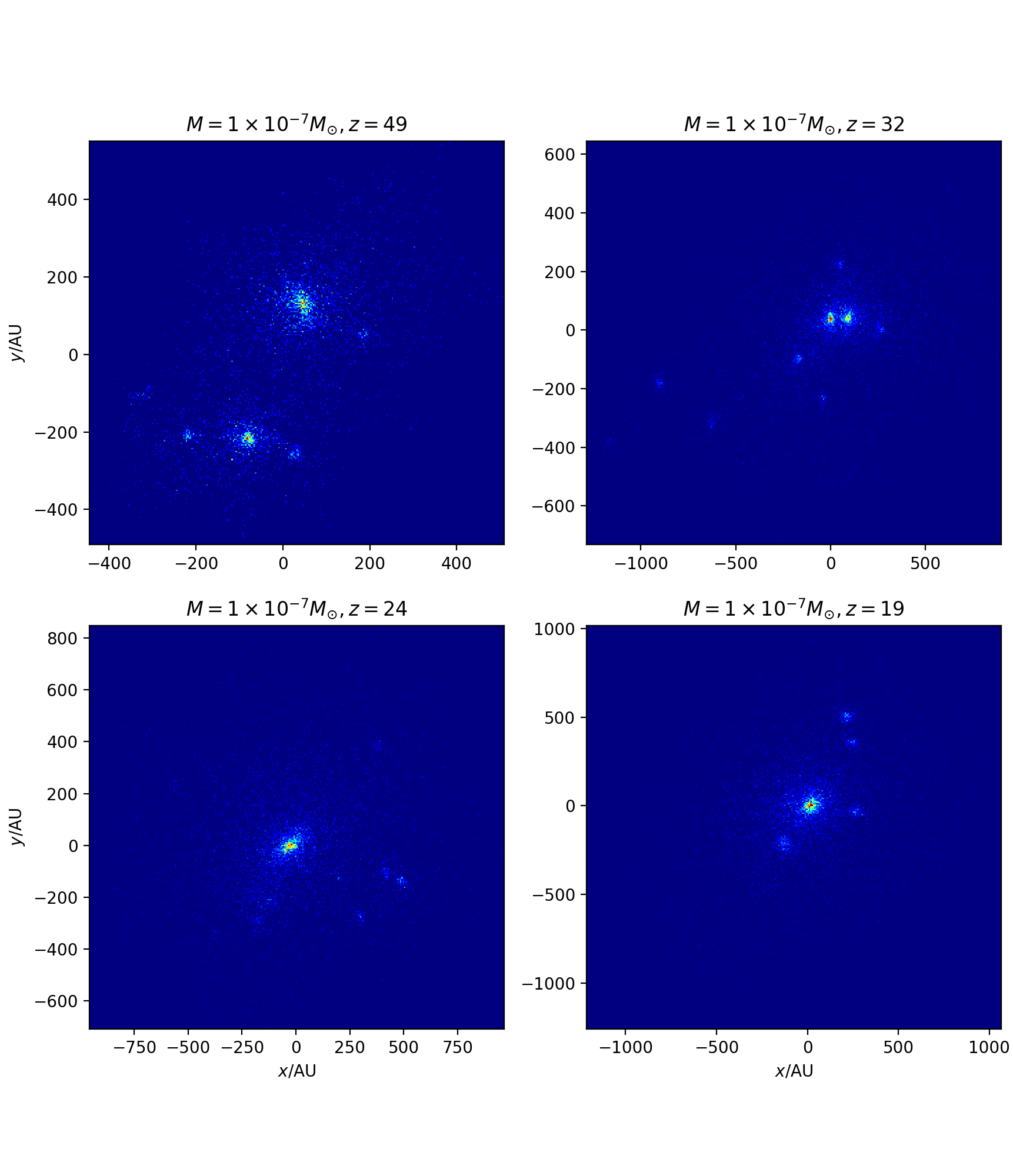}
\end{minipage}
\hfill
\begin{minipage}[b]{0.49\textwidth}
\includegraphics[width= 1.02\textwidth]{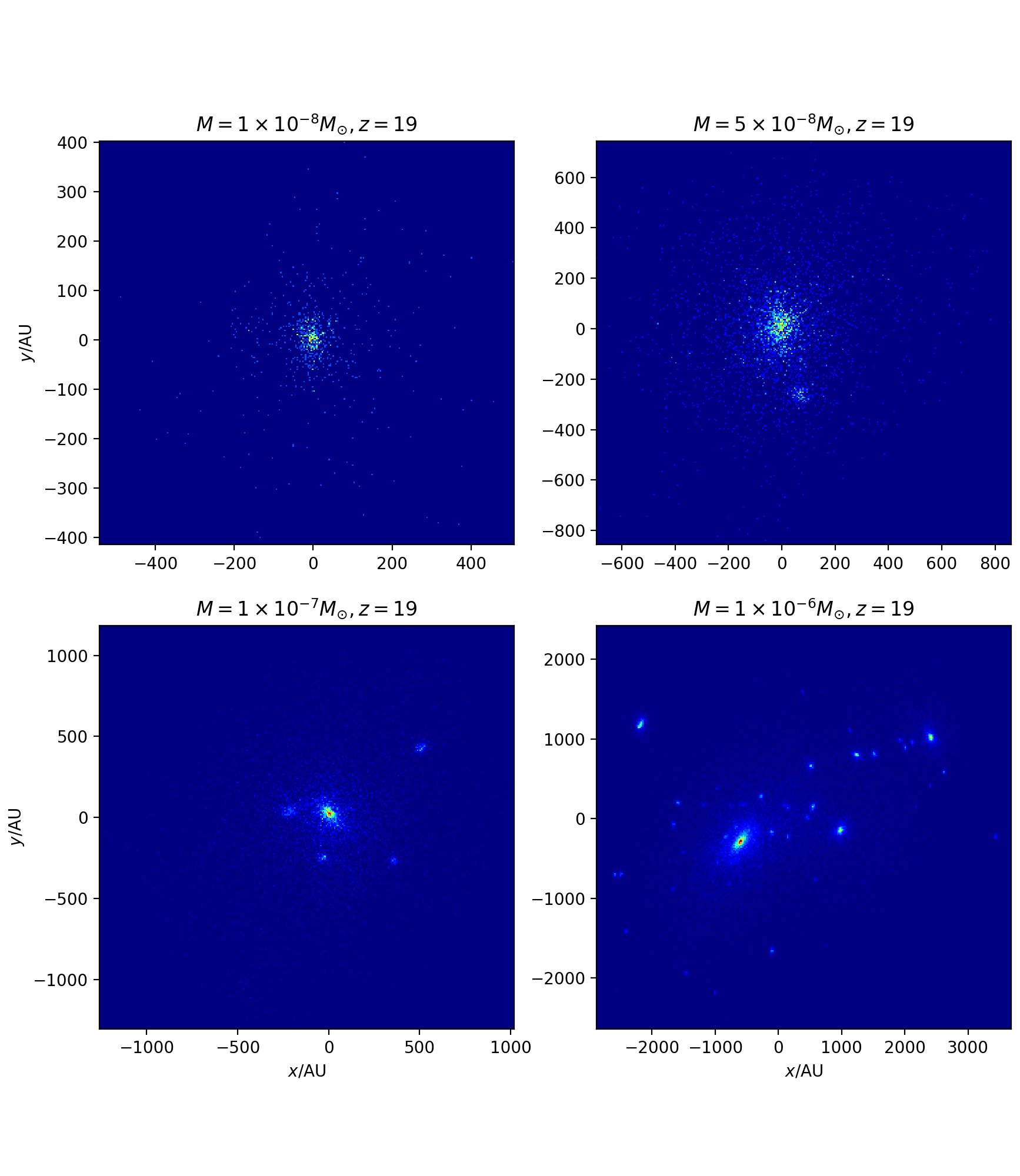}
\end{minipage}
\caption{The visualization of halos that have different masses and redshifts from our simulation interpreted with $A_{\rm osc}=0.1$ and $k_{\rm osc}=19.8~ \rm pc^{-1}$. Left panels are showing halos with the same mass $10^{-7}M_{\odot}$ at different redshifts. Right panels are showing halos with different masses at the same redshift. 
The density profile is supposed to be the same for halos that have the same masses. However, we show that the halo at $z=49$ has more substructures than other halos with the same mass. This is because $M=10^{-7}M_{\odot}$ is at the high mass end of the mass function at $z=49$ and the high mass end shifts to larger masses at lower redshifts. Halos at high mass end just formed very recently and did not go through enough merger events. Therefore they may have a lot of substructures than other halos. In summary, a larger value of $\nu = \delta_c/\sigma(M,z)$ is corresponding to more substructures.}
\label{fig:halo_vis}
\end{figure*}

We plot NFW density profiles at z=19 and z=49 in Fig.~\ref{fig:density_profile_z=19} and Fig.~\ref{fig:density_profile_z=49} respectively and compare them with the density profiles of 10 halos taken from the simulation that fall nearest the specified mass. The NFW profile is fully determined by the halo mass $M$ and scale radius in Eq.~\ref{eq:scale_radius} as $\rho_s = {M}/[4\pi r_s^3({\rm log}(1+c)-c/(1+c))]$, while the colored curves are computed from the simulation data, representing density profiles of 10 halos around the same halo mass $M$. Many of the density profiles of small halos agree with solid NFW profile reasonably. However, some of the halo density profiles do not agree with the NFW profile, with a flatter density profile in the central region.  This occurs more often for our most massive halos ($M\sim 10^{-6}M_{\odot}$).  We do not expect this to be a gravitational softening effect because the flat density profile starts on a scale much larger than the softening scale (and is most common in our massive halos). Rather, it owes to halos with multiple components that have not merged:  The largest minihalos -- like galaxy clusters today -- are the least merged and relaxed systems.  The large amount of structure in our massive halos can be seen in Fig.~\ref{fig:halo_vis}, which plots the projected density distribution of four $M\sim 10^{-7} M_\odot$ halos at different redshifts and four different mass halos at $z=19$. For instance, the distance between the substructures for the axion minihalo with $M=10^{-7}M_{\odot}$ at $z=49$ is roughly 200 AU ($10^{-3}\rm pc$) as shown in Fig.~\ref{fig:halo_vis}, which matches with the size of the flat central region of high mass halos in Fig.~\ref{fig:density_profile_z=49}.
More generally, in our white noise cosmology, halo formation for a given range of halo masses is spread over a larger range in scale factor than in the standard cosmology. This enhances the density contrast of substructures and makes them more able to survive in their parent halo.

In the picture that the axion halo forms early and then sits undisturbed, one might expect the characteristic scale radius, $r_s$, is only a function of halo mass $M$ and does not depend on redshift \cite{Dai:2019lud}. (The virial radius as defined, however, does depend on the redshift because it is determined by the mean density of the Universe. Therefore, $c$ grows linearly with scale factor $a$ in this picture.) Dai \& Miralda-Escud\'e \cite{Dai:2019lud} assumed that the axion minihalos form with a relatively small concentration of $c=4$ and that the collapse at a given minihalo mass that can be characterized by when a $1\sigma$ perturbation at that mass collapses. These assumptions are sufficient to predict $r_s$ and $c$ analytically as a function of halo mass \cite{Dai:2019lud}. 
If we calibrate the analytical prediction with a prefactor and compare it to the average density profiles in the simulation, we found they are in good agreement for different halo masses at different redshifts. Therefore, this picture seems to be roughly obeyed by our simulations. The solid curves in Fig.~\ref{fig:density_profile_z=19} and Fig.~\ref{fig:density_profile_z=49} show
\begin{equation}\label{eq:scale_radius}
\begin{split}
    r_s(M)\approx &3.7\times 10^{-3} h^{-1}{\rm pc}\left(\frac{A_{\rm osc}M_0}{10^{-11}M_{\odot}/h}\right)^{-1/2}\\
    &\times \left(\frac{M}{10^{-6}M_{\odot}/h}\right)^{5/6},
\end{split}
\end{equation}
or equivalently
\begin{equation}\label{eq:mass_concentration_relation}
    c(z) \equiv \frac{r_{\rm vir}}{r_s} = \frac{1.4\times 10^{4}}{(1+z)\sqrt{M/(A_{\rm osc}M_0)}},
\end{equation}
where $M_0$ is the characteristic mass of axion minihalos at power spectrum cutoff, which is defined in Eq.~\ref{eq:M0}, and $A_{\rm osc}$ is the amplitude in Eq.~\ref{eq:power}.   The only relevant parameter in our simulation is $A_{\rm osc}/k_{\rm osc}^3$, which is proportional to $A_{\rm osc}M_0$.  The dependencies of Eqns.~\ref{eq:scale_radius} and \ref{eq:mass_concentration_relation} are from the model of \cite{Dai:2019lud}, discussed shortly. In Ref. \cite{Eggemeier:2019khm}, characteristic concentration parameters for axion minihalos at $z=99$ for certain mass ranges are discussed.  
Their concentration parameter varies with mass more slowly than our relation at the low mass end, which is still reasonably consistent with our model because their power spectrum is also more flat at smaller scales. At highest masses they study of $\sim 10^{-8}M_{\odot}$, their concentration parameters roughly agree with what we found.

Figures~\ref{fig:density_profile_z=19} and \ref{fig:density_profile_z=49} show respectively $z=19$ and $z=49$ four four halo mass bins that span $2-3$ orders of magnitude in mass.  The solid curve is the $r_s(M)$ given by Eq.~\ref{eq:scale_radius}, and the dashed is a factor of two smaller.  The former values of $r_s$ and $c$ appear to describe most of the relaxed halo profiles, and the outer profile of of the unrelaxed. 

We also note that these concentrations are much larger and have a stronger mass dependence than standard CDM halos \citep{2017arXiv171105277O}.  We expect CDM halos only to have a similar concentration if they form at the same redshift as our minihalos.  Finally, it is well known that CDM halos have a lognormal distribution of concentrations with a full width at half maximum (FWHM) of $\approx 1~$dex at $z\approx 0$. For our relaxed axion minihalos, there is no clear evidence for such significant scatter in their concentrations although we did not study it quantitatively.

Our simulated minihalos halos have $r_s$ values that are smaller than the analytic prediction in \cite{Dai:2019lud} by a factor of four. We think that some of this difference comes from the fact that they defined the halo mass when its concentration number is four. The scale radius of a halo will remain a constant as time evolves, but the `boundary' at $r_{\rm vir}$ will expand due to the less dense Universe. Therefore, dark matter halos will go through `pseudo-growth' as the boundary expands and encompasses more mass \cite{2013ApJ...766...25D}.  For the NFW profile that describes halos in our simulations, a halo with $c=100$ at late times is about four times heavier than it was at formation with $c=4$ simply owing to this effect. 

\section{axion minihalo observables}
\label{sec:observables} \label{sec:observations}

\begin{figure}
\includegraphics[width= 0.47\textwidth]{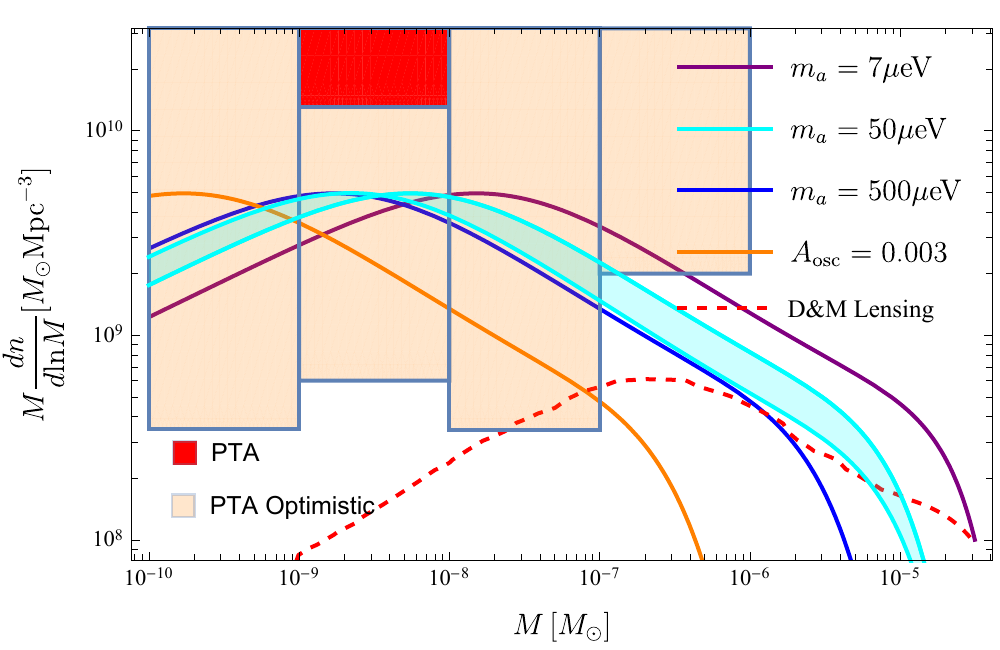}
\caption{Model prediction of axion halo mass function at the present time with different choice of axion masses (solid curves) compared against forecasts for the observational sensitivity for a future pulsar timing array (PTA) observations and gravitational lensing in the scenario proposed in \citet{Dai:2019lud}. Disruption is neglected for reasons discussed in \S~\ref{sec:disruption}. Different axion particle masses are shown computed assuming $A_{\rm osc} = 0.03$ for the power spectrum normalization, and our amplitude parameter can be deduced from the masses $m_a$ via Eq.~\ref{eq:axion_mass} and we show one string-dominated case with $A_{\rm osc} = 0.003$ and $m_a = 500\mu eV$ (orange curve). The power spectrum with mass $m_a=50\mu \rm eV$ and $A_{\rm osc} = 0.03$ agrees with the white scaling of the early universe axion simulations used in \cite{Eggemeier:2019khm}. We show with the yellow band for our middle-of-the road $50\; \mu$eV mass the range of estimates using the turnaround and collapse $\delta_c$ for when axion minihalo growth is terminated. 
The red region represents the PTA sensitivity with SKA parameters, while the orange region assumes more optimistic PTA parameters (see main text).
The dashed curve presents the mass function in Dai et al. \cite{Dai:2019lud}, which owing to the size of an effect it creates on this observable, we interpret as (very roughly) a lower bound on what can be probed by lensing of magnified stars (see text). }
\label{fig:current_mf}
\end{figure}

The previous sections obtained the mass function and density profile of axion minihalos from simulations and developed a semi-analytic model to extrapolate to the present day. We can now use this model to evaluate the prospects for detecting axion minihalos in the post-inflation scenario. We will show that pulsar timing arrays and the gravitational lensing of highly magnified stars both show promise for constraining our mass spectrum of axion minihalos. 

Figure \ref{fig:current_mf} is the key figure on which we will place our constraints.  It shows our semi-analytic model's prediction for the present time axion minihalo mass function compared to forecasts for the observational sensitivity for future pulsar timing array (PTA) observations and the gravitational lensing in the scenario proposed in \citet{Dai:2019lud}. Different axion masses are shown assuming $A_{\rm osc} = 0.03$ (and for one curve $A_{\rm osc}= 0.03$) for the power spectrum normalization.  The power spectrum amplitude $A_{\rm osc}/k_{\rm osc}^3 $ can be deduced from the masses $m_A$ via Eq.~\ref{eq:axion_mass}.  The range of axion masses reflects the uncertainty in how much axion strings contribute to the production of early universe axions.  The blue and solid curves take $m_A=500\mu eV$ and represent the scenario of axion strings radiating mostly in the infrared and dominating over misalignment production by the factor $\log(f_A/m_A)\sim 60$.  We note that $A_{\rm osc}= 0.03$ is calibrated to match the low wavenumbers of the simulations of \citet{Vaquero:2018tib} where strings provide a middle-of-the-road enhancement.  Pure misalignment production of axions likely has a larger value for $A_{\rm osc}$. However, even a string dominated spectrum should have a significant $A_{\rm osc}$ owing to the expectation that there are a handful of strings per Hubble patch. 

We now compare these model predictions to the forecast sensitivities of pulsar time arrays and gravitational lensing observables.

\subsection{Pulsar Timing Array}
Due to the stability of pulsar pulse phases observed over at least several year durations, dark matter structures around the Earth-pulsar system can imprint discernible signatures in these phases via gravitational Doppler and Shapiro delays. Using pulsar timing arrays (PTAs), individual transiting subhalos can be detected in the future by Square Kilometer Array (SKA) \cite{Rosado:2015epa}. Constraints for a single `subhalo' mass for dark matter substructure are presented in \cite{Ramani:2020hdo}, which \cite{Carr:2017jsz} found can be extended to non-monochromatic mass functions within a mass bin of $M_1 < M < M_2$ using the following inequality:
\begin{equation}\label{eq:PTA_sens}
    \int_{M_1}^{M_2} dM \frac{dn}{d{\rm ln}M}\frac{1}{\bar \rho_{\rm dm}f_{\rm max}(M, c)}\le 1.
\end{equation}
Here, $f_{\rm max}$ is the maximum mass fraction of dark matter in substructures of a \emph{single} mass presented in \cite{Ramani:2020hdo}, which is a function of halo mass and concentration number, $\bar \rho_{\rm dm}$ is the mean density of dark matter in the Universe, and $dn/d{\rm ln}M$ is the mass function of axion minihalos. We consider the two observational scenarios for $f_{\rm max}$ presented in \cite{Ramani:2020hdo}.  One is an optimistic case based on the futuristic PTA parameters $N_P=1000,\, T=30 {\rm\, yr},\,t_{\rm rms}=10 \rm\, ns,\, \Delta t= {1\rm \,week}$, where $N_P$ is the number of pulsars, $T$ is the observing duration, $t_{\rm rms}$ is the residual timing noise and $\Delta t$ is the cadence.  The second is a somewhat less futuristic PTA sensitivities based on the estimated capability of the PTA with the Square Kilometer Array: $N_P=200,\, T=20 {\rm\, yr},\,t_{\rm rms}=50 \rm\, ns,\, \Delta t= {2\rm \,weeks}$. The parameter choices in both cases are motivated further in \cite{Ramani:2020hdo}. 

With our mass function and concentration numbers of axion minihalos at present day, we use Eq.~\ref{eq:PTA_sens} to determine whether PTA observations with future instruments may be sensitive to axion minihalos. We present our results in Fig.~\ref{fig:current_mf}, where the current mass function of axion minihalos and the threshold mass function that will lead to a detectable signal in PTA observations are plotted.  As shown in Fig.~\ref{fig:current_mf}, our estimate for the current mass function of axion halos lives well above the threshold that would be detectable for future PTA observations with optimistic PTA parameters. This conclusion holds for each of the axion mass scenarios we consider. However, our estimate for the current mass function is not large enough to produce a detectable in PTA observations with SKA parameters. We need more futuristic observations to detect axion minihalos with PTAs. (See this footnote for a caveat on our PTA bounds on the most massive halos that pertains to their lower concentrations: \footnote{The PTA bounds in Fig.~\ref{fig:current_mf} assume $c\gtrsim 1000$, which we find applies at $10^{-8}M_{\odot}$ (Eq.~\ref{eq:mass_concentration_relation}). When the halo mass is larger than $10^{-7}M_{\odot}$, axion minihalos have lower concentration numbers and so the bound shown in Figure~\ref{eq:PTA_sens} are not applicable and a more detailed study is required.})

In \S~\ref{sec:disruption} we consider disruption by encounters with stars.  We 
 argue that this process is not efficient enough to suppress by the order of magnitude required to be below the `PTA Optimistic' limits. 

\subsection{Gravitational Lensing of highly magnified stars }
The usual microlensing signatures of axion minihalos requires the concentration number to be ultrahigh in order to exceed the critical surface density for lensing ($c \gtrsim10^{7}$ for typical masses).  Our minihalos do not come anywhere near such high concentrations (Eq.~\ref{eq:mass_concentration_relation}). However, recently another lensing diagnostic was suggested. Caustic transiting stars behind a galaxy cluster lens can reach extreme magnifications of $\mu \gtrsim 10^{3}\operatorname{--} 10^{4}$ as the lensed stars cross microlensing caustics induced by intracluster stars \citep{2018ApJ...857...25D, 2017ApJ...850...49V}.  The perturbations from dark matter structure can add additional structure to these lensing caustics, changing their profile \cite{Dai:2019lud}.  Indeed, \cite{Dai:2019lud} argued that this effect is sensitive to the minihalos in the post-inflationary axion scenario.

In particular, Dai et al. \cite{Dai:2019lud} showed that this diagnostic is sensitive to convergence fluctuations are at the level of $\Delta_{\kappa}\sim 10^{-4}\operatorname{--}10^{-3}$ on scales $10\operatorname{--}10^{4}{\rm AU}/h$ (corresponding to mass scales $10^{-8}\operatorname{--}10^{-5}M_{\odot}$).  Motivated by the fact that axion minihalos have $r_s$ in this interesting range of scales, they modeled the abundance of axions using the $z\sim 1$ Press-Schechter mass function calculated with both axion isocurvature and inflationary adiabatic fluctuations. (This is likely a conservative model as it predicts most of the mass is in large halos that form from the adiabatic fluctuations. Others have also adopted this model, which we contrast with in the conclusions.)  They concluded that the post-inflation axion scenario can produce sufficient level of fluctuations, although their results suggest that their QCD axion model is near the minimum of what might feasibly be detected. Our simulation results suggest a larger amplitude for the mass function at all masses except the scenario where axion mass is $500\mu \rm eV$, indicating that axion minihalos likely lead to a larger lensing signature than in their fiducial model. In Fig.~\ref{fig:current_mf}, we show the comparison between the mass function used for the lensing calculations in \cite{Dai:2019lud} (dashed red curve) against our mass functions.

We also find in \S~\ref{sec:disruption} that stellar disruption is unlikely to destroy the axion minihalos in the cluster environment envisioned in \cite{Dai:2019lud}, verifying simpler estimates they present.  This may not be the case for a similar diagnostic involving a galactic macro-lens rather than a cluster.

\section{disruption of axion minihalos}
\label{sec:disruption}
\begin{figure*}
\includegraphics[width= 1\textwidth]{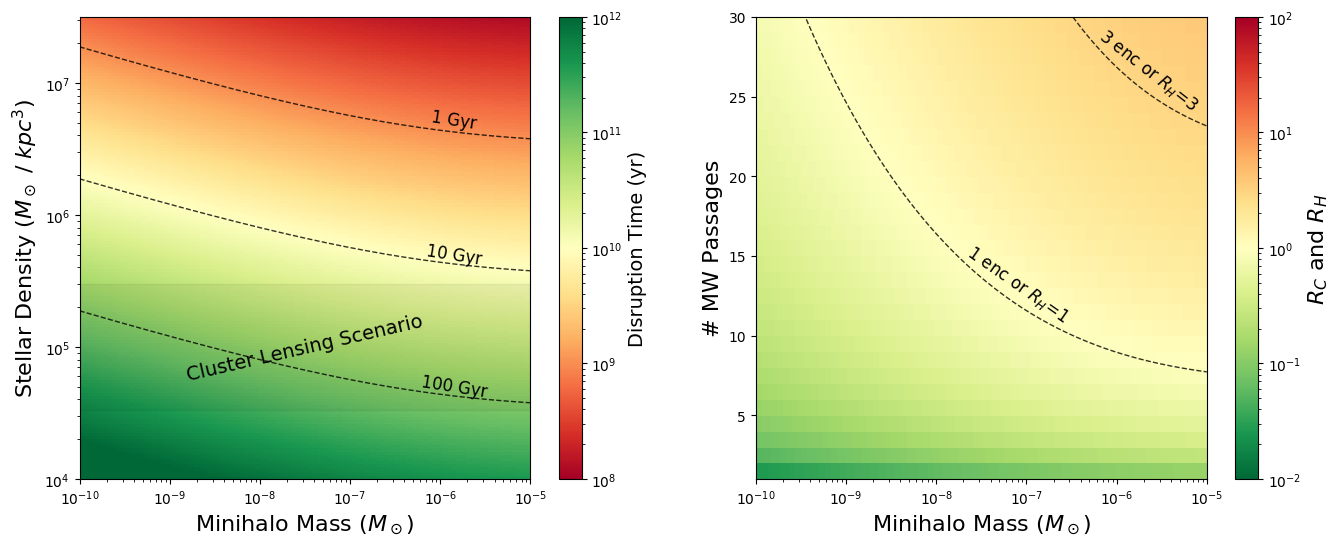}
\caption{
Disruption of axion minihalos from stellar encounters, assuming NFW density profiles with concentrations calculated for $A_{\rm osc}=0.03$ and $m_a=50\mu$eV. The left panel shows the average time taken for one catastrophic ($\Delta E > |E_b|$) encounter to occur, which is also the approximate time taken for dynamical heating from smaller encounters to add up to $|E_b|$. Most minihalo should survive in the environment needed for the gravitational lensing observable. The right panel shows both disruption effects as a function of passages through the Milky Way disk. 
}
\label{fig:Minihalo_Disruption}
\end{figure*}


The primary mechanism for disrupting minihalos is high speed encounters with stars and other minihalos. Here we present estimates of the disruption from these encounters. We use the NFW profiles discussed in \S~\ref{sec:density_profile}, and the impulsive analytic framework developed in \citet{Spitzer1958, Gerard1983, Moore1993, Carr1999, Binney2008, Green_2007}. The associated energy change from an encounter is \citep{Green_2007}:
\begin{equation} \label{eq:enc_energy}
    \Delta E = \frac{4\alpha^2}{3} \frac{G^2M_p^2 M r_{\rm vir}^2}{V^2b^4},
\end{equation}
where $M_p$ is the perturber mass, $b$ is the impact parameter, $M$ is the subject axion minihalo mass, 
and $\alpha^2$ is the root mean square radius ($0.1 \lesssim \alpha^2 \lesssim 0.4$; the exact expression can be calculated from the definitions in \citep{Green_2007}). 

Once minihalos fall onto a CDM halo, we assume they stop accreting mass and evolving in any significant way. Thus, for the purposes of our disruption analysis, we evaluate $r_{\rm vir}=c r_s$ as a static quantity at $z=20$ using Eq.~\ref{eq:mass_concentration_relation}.  This has the effect of cutting off the NFW halo at the $z=20$ virial radius, which makes the axions more bound than a NFW halo that extends out to the $z=0$ virial radius. 

To consider the disruption from an encounter, we compare the energy imparted to the binding energy of the subject minihalo: $|E_b| = f_c G M^2 / 2 r_{\rm vir}$, where $f_c$ is computed for an NFW profile and ranges in value from 0.75 for $10^{-4}M_\odot$ to 12 for $10^{-10}M_\odot$ minihalos using our fitted value for the concentration. There are two important limits: 1) catastrophic encounters where $\Delta E / |E_b| > 1$, and 2) dynamical heating from the cumulative effect of many small encounters. The catastrophic regime occurs at impact parameters satisfying this condition, the largest of which we define to be $b_c$. 

Over its lifespan, a minihalo will have many encounters with other minihalos and stars. Minihalo-minihalo interactions occur in a larger macro halo, and are negligible in galactic halos or cluster halos, where high encounter speeds and low perturber masses prevent catastrophic encounters entirely and suppress the heating rates. However, in the early stages of structure formation the minihalos inhabit halos with velocities not much larger than their virial velocities, and so these effects could be important.  While preliminary calculations suggest this process is less important than stellar disruptions (aided by fact that often axion minihalos accrete onto much larger CDM halos), we plan further study of minihalo interactions in future work. 

We now focus on stellar encounters. Assuming Maxwellian velocity distributions, an estimate for the catastrophic encounter rate is (Appendix~\ref{appendix:disruption}):
\begin{equation}\label{eq:enc_catastrophic_rate}
    R_c = \sqrt{\frac{8G\alpha^2}{3f_c M}}\pi \rho_p r_{\rm vir}^{3/2}.
\end{equation}
This rate does not depend on the velocity dispersion nor the distribution of perturbers masses, just the overall perturber mass density, $\rho_p \equiv M_p n_p$, and subject mass. A power law fitting can approximate this result when $M<10^{-5}M_\odot$:
\begin{equation}\label{eq:enc_catastrophic_dis_time_approx}
    \begin{split}
    R_{\rm c}^{-1} \approx 40\ \mathrm{ Gyr} \bigg(\frac{\rho_p}{10^{5} M_\odot \mathrm{kpc}^{-3} }\bigg)^{-1}  \\
    \bigg(\frac{M}{10^{-6}M_\odot} \bigg)^{-0.16} \bigg( \frac{A_{\rm osc}}{0.03} \bigg)^{0.11},
    \end{split}
\end{equation}
The $M$ and $A_{\rm osc}$ dependence is weak.

The next consideration is the cumulative heating of smaller encounters, with $b>b_c$. Successive smaller encounters continue to add energy, leading to disruption when the total energy added is $\sim |E_b|$. We estimate that the heating rate is (Appendix~\ref{appendix:disruption}): 
\begin{equation}\label{eq:heating_rate}
    \begin{split}
    R_H \equiv \frac{\dot{E}}{|E_b|} = \sqrt{\frac{8\alpha^2G}{3f_c M}}\pi \rho_p r_{\rm vir}^{3/2} =  R_c. 
    \end{split}
\end{equation}

Curiously our estimate for the heating rate is equal to our estimate for the catastrophic destruction rate, Eq.~\ref{eq:enc_catastrophic_rate}. The average time for one catastrophic encounter to occur is also the average time needed for cumulative heating to add up to the binding energy. Details on this interesting equivalence can be found in Appendix~\ref{appendix:disruption}. 

At small impact parameters of $b \lesssim r_s$, Eq.~\ref{eq:enc_energy} overpredicts the energy imparted \citep{Carr1999, Green_2007}. Simulations exploring this in \citep{Green_2007} determined that the transition is well-fit by a sharp cutoff between these two regimes, located at $b_s \equiv(4\alpha^2/9\beta^2)^{1/4}r_{\rm vir}$, where $\beta^2$ is the is the root mean square inverse radius and is $6 < \beta^2 < 14$ for our minihalos (see \citep{Green_2007} for details). We find that, for the minihalos in this work, $b_s=[0.25-0.41]r_{\rm vir}$. 
  For encounters with $b_s>b_c$, catastrophic encounters are impossible.  We calculate $b_c$ and $b_s$ and find that, for speeds characteristic of the cluster lensing scenario ($V=1500$ km s$^{-1}$), minihalo of mass $10^{-5} M_\odot$ or larger have $b_s > b_c$ for encounters with stars as high as $1.6M_\odot$. This implies that the largest minihalos are immune to the majority of would-be-catastrophic encounters in this environment; the encounters contribute to heating instead. We neglect this in the following computations and note that the net effect is longer disruption times for $\gtrsim 10^{-5} M_\odot$ minihalos in galaxy clusters.  
  This consideration is even less relevant in Milky Way-like environments, 
  where the vast majority of minihalos predicted by our simulations are nearly unaffected by this correction.


Using $R_H$ and $R_c$, we estimate the disruption of axion minihalos in environments relevant for the Milky Way and cluster environments. For the Milky Way, we calculate the expected catastrophic encounters and total heating in terms of the number of passages through the Milky Way's disk.  This parameterization allows for the fact that some minihalos will have encountered the disk a handful of times if they were accreted late and had a low angular momentum orbit that takes it far out into the halo.  Other minihalos at the Solar Circle likely pass through the disk many tens of times.   We neglect stars in the bulge and stellar halo, estimate the stellar surface density at 8 kpc to be $10^8 M_\odot$ kpc$^{-2}$ and $V=200$ km/s.  Results are shown in right panel of Figure~\ref{fig:Minihalo_Disruption}.

In the Milky Way, relevant to the PTA observable, our estimates show that $\sim 10^{-6}M_\odot$ minihalos experience a catastrophic encounter and disruptive heating after $\sim10$ disk passages, with stellar impacts becoming somewhat less important for less massive minihalos. The minihalos accreted many Gyr ago likely would have had 1-3 catastrophic encounters, and those that avoided one would have had significant heating.  Material that has fallen onto the Milky Way more recently or that is on an orbit with rare disk passages is likely to retain its axion minihalos, even the larger ones.  More detailed calculations are required to make a better estimate. Our results suggest disruption patterns similar to \citet{Kavanagh2020}, despite differences in the modeling of the minihalos.

Detecting the perturbations from axion minihalos from gravitational lensing of highly magnified stars (\S \ref{sec:observables}) depends on whether minihalos are disrupted in the centers of galaxy clusters (although minihalos outside of the cluster can contribute non-negligibly).  Galaxy clusters are a less stellar rich location than the Milky Way and, thus, survival is more likely. We calculate the expected disruption time for this cluster environment; results are shown in the left panel of Figure \ref{fig:Minihalo_Disruption}.  At these lower densities, our estimates indicate that many minihalo do not experience a single catastrophic encounter and collective heating should not significantly alter the axion minihalos, in agreement with the estimates in \citet{Dai:2019lud}.

Authors studying the effects of these disruptive mechanisms on dark halos and stellar systems have made the case that the result from $\Delta E > E_b$ encounters is not full disruption, but mass loss \cite{Gieles2016, Kavanagh2020, vandenBosch2017, Delos2019}. The energy imparted from a catastrophic encounter is mostly carried off by particles in outer layers, leaving behind dense cores.  More detailed simulations would be required to understand the masses of these cores for the minihalos that are most impacted by encounters.

\section{Conclusions}
We have run N-body simulations that study the formation of axion minihalos on mass scales where the initial spectrum of axion perturbations is described by Gaussian white-noise. The exact spectrum of axion perturbations will be damped below the white noise scaling at high momentum, and high-wavenumber modes can be shaped by non-gaussianities.  However, lower momenta modes where the spectrum is white and that limit to being Gaussian are likely to be most relevant for observations, as they shape the late-time minihalo mass function.

We simulated the mass function and density profiles of axion minihalos. Our results show that the standard Press-Schechter and Sheth-Tormen mass functions only err at the factor of $\sim 2$ level.  We further showed that a tweaked version of the Sheth-Tormen mass function and the NFW profile with a physically motivated scale radius (adapting the model of \citet{Dai:2019lud}) accurately describe the simulated halo properties at different redshifts.  We further showed that our mass function, developed on our Gaussian white noise simulations, is also able to roughly describe the evolution of the mass function found in the simulations of \citet{Eggemeier:2019jsu} that start from the outputs of early universe simulations to the sine-Gordon equation.  

Our cosmologically minute simulations do not capture the range of structures that collapse in the real universe.  We developed a model which allows us to extrapolate our results to the current time and make connection to observations. Namely, we assume that axion minihalos  stop merging and growing once they fall into the much larger halos sourced by the adiabatic fluctuations from inflation, but that they are able to survive intact. 
This extension of our model allowed us to make predictions for pulsar timing arrays and microlensing of highly magnified cosmological stars. The abundance of axion minihalos we find appear to be above the threshold that can be probed by PTA observation with futuristic parameters, modulo uncertainty in the amplitude of the density fluctuations when cosmic strings dominate the axion production. Our predictions for halo mass function and concentration are also greater than an estimate for the threshold sensitivity of the microlensing scenario. 

However, these predictions ignore processes that potentially disrupt axion minihalos within larger dark matter halos.
We investigated the disruption of minihalos by stellar encounters. While we found that such disruption is unlikely to be important within galaxy clusters (applicable for the microlensing observable), we found that it is more important in a galactic environment relevant to PTAs; however, they are likely not disruptive enough to suppress our predictions by more than an ${\cal O}(1)$ factor.  Semi-analytic techniques that better approximate stellar encounters as well as other disruptions processes, such as tides from the parent halo, have been developed \citep{2001ApJ...559..716T, Green_2007, 2020arXiv201107077E}.  We aim to apply these to improve our estimates for the abundance of axion minihalos in future work.  In lieu of these calculations, our primary result for the mass function (without disruption) in Fig.~\ref{fig:current_mf} should perhaps be considered an upper bound for the axion minihalo abundance.  To further complicate the story, axion minihalos have more substructure than CDM halos, which is not included in our model and would enhance the signals.

Our estimates for the abundance of minihalos in present-day systems contrast with previous studies \cite{Dai:2019lud,Lee:2020wfn} where minihalos that merged into other halos before they were subsumed into the Milky Way or galaxy cluster halos were effectively excluded. In particular, \cite{Dai:2019lud,Lee:2020wfn} computed the mass function results from axion isocurvature plus adiabatic fluctuations at the time of formation of the Milky Way or cluster, which results in the mass function of axion halos being shifted to considerably higher masses and less concentrated halos compared to our estimates (compare the dashed `D\&M Lensing' curve in Fig~\ref{fig:current_mf} with the solid curves).  Our estimates include the less massive more concentrated minihalos that merge earlier with CDM halos.  We argued that many of these should survive.

Our simulations are scale-invariant except on the smallest scales, and most collapse occurs during matter domination when the evolution is self-similar.  In these limits, the only relevant parameter is the variance on any mass scale $\sigma^2(M)$. 
 Our semi-analytic halo mass function naturally incorporates this symmetry.
Thus, it is straightforward to apply our results to other axion-like particles (ALP), where the axion decay constant $f_a$ and axion mass $m_a$ are two potentially independent model parameters unlike for the QCD axion. 
  Our simulations apply to the white noise formation for all ALP scenarios by only matching the mass scale that has the same variance as in our calculations.
 Disruption processes do depend on halo mass and so unfortunately our results there do not generalize.  Of course, the observables will change with mass as well. For example, much more massive halos may exceed to the critical density to act as a strong gravitational lens \citep{PhysRevD.97.083502}. 

Our calculations also have relevance to much different dark matter scenarios.  Primordial black holes (PBHs) should have a white spectrum much like the post-inflation axion, with ${\cal O}(1)$ fluctuations on the mass scale of the black holes, $M_{\rm PBH}$ \cite{2019PhRvD.100h3528I}.  Our simulations and calculations also apply to the case where these comprise the dark matter for halos comprised of PBHs with $M\gg M_{\rm PBH}$ (although the center of PBH halos could be cored by two body interactions).  Indeed, there is still a window at $M_{\rm PBH} = 10^{-17} -10^{-11} M_\odot$ where PBHs could be a substantial fraction of the dark matter \citep[e.g.][]{2019JCAP...08..031M}, one of the few windows left. Our calculations suggest that microlensing and PTAs may be able to constrain the more massive end of this window.  Finally, our white spectrum of perturbations is closer to the blue spectrum anticipated from scenarios with early matter domination \citep{2015PhRvD..92j3505E} or a light vector whose abundance is set by inflationary fluctuations \citep{2016PhRvD..93j3520G}, suggesting that the standard semi-analytic mass functions will apply there as well.

\acknowledgements
We thank Vid Ir\v si\v c, Liang Dai, Simeon Bird, Kathryn Zurek, Stephen Sharpe and Masha Baryakhtar for useful conversations. HX would like to thank Simeon Bird for his help on running MP-Gadget. We thank Liang Dai for useful comments on primordial black holes.
This work was supported by the University of Washington Royalty Research Fund. HX is also supported
in part by the U.S. Department of Energy under grant number DE-SC0011637 and the Kenneth
K. Young Chair in Physics.
This work used the Extreme Science and Engineering Discovery Environment (XSEDE) \cite{xsede,ecss}, which is supported by National Science Foundation grant number ACI-1548562.

\appendix
    \section{Fitting the Halo Mass Function}\label{appendix:Fitting}
The halo mass function from our simulation does not agree with the prediction by Press $\&$ Schechter \cite{1974ApJ...187..425P} or Sheth $\&$ Tormen \cite{1999MNRAS.308..119S} computed with power spectrum in Eq.\ref{eq:power}. However, we can calibrate our mass function with these analytic models and extrapolate the result of our simulation to present time.
A simple model of collapsed halos by Press $\&$ Schechter \cite{1974ApJ...187..425P} gives the following halo mass function:
\begin{equation}
    \frac{m^2dn/dm}{\bar{\rho}}\frac{dm}{m}=\nu f(\nu)\frac{d\nu}{\nu},
\end{equation}
where $\bar{\rho}$ is the comoving density of matter and $\nu, f(\nu)$ are defined as:
\begin{equation}
\begin{split}
    &\nu f(\nu)=\sqrt{\frac{\nu}{2\pi}}{\rm exp}(-\nu/2),\\
    &\nu \equiv \frac{\delta_c^2(z)}{\sigma^2(m)},\\
\end{split}
\end{equation}
where $\delta_c$ is the critical density required for spherical collapse at z. In an Einstein-de Sitter cosmology, $\delta_c=1.686$. $\sigma^2(m)$ is the variance in the initial density fluctuation field when smoothed with a tophat filter of scale $R=(3m/4\pi\bar{\rho})^{1/3}$, which can be determined as:
\begin{equation}
\sigma^2(m)\equiv \int \frac{dk}{k}\frac{k^3P(k)}{2\pi^2}\vert W(kR)\vert^2,
\end{equation}
where $W(x) = (3/x^3)[{\rm sin}(x)-x{\rm cos}(x)]$ 
is the spherical top-hat window function. The variance of white-noise power spectrum from axion can be expressed as:
\begin{equation}
   \sigma(M)=\sqrt{\frac{3A_{\rm osc}}{2\pi^2}\frac{M_0}{M}}. 
\end{equation}

The Sheth-Tormen mass function provides a better fit to the number density of halos in simulations, which gives:
\begin{equation}
    \nu f(\nu)=A(p)\left(1+(q\nu)^{-p}\right)\left(\frac{q\nu}{2\pi}\right)^{1/2}{\rm exp}(-q\nu/2),
\end{equation}
where $p\approx 0.3$, $q\approx0.75$ and  $A(p)=[1+2^{-p}\Gamma(1/2-p)/\sqrt{\pi}]^{-1}\approx 0.3222$ \citep{1999MNRAS.308..119S}.  The functional form of this mass function is motivated by the impact of large-scale tidal fields on delaying collapse relative to the (spherical) Press-Schechter prediction.

We can fit the halo mass function from our simulation by tuning the parameters $A,~p,~q$ in the Sheth-Tormen mass function.  We treat model parameter $A$ as a free parameter rather than expressing it in terms of $p$. We find the best fitting parameters to be $A=0.374,p=0.19,q=1.2$. Given these parameters, we can make predictions in halo mass function at different redshifts. As shown in Fig.~\ref{fig:mass_function}, the fitted halo mass function broadly agrees with the mass function obtained from simulation, except at low mass end where halo mass $M\lesssim 10^{-9}M_{\odot}$.  See the main text for discussion of the differences at low masses.


\section{Convergence testing}\label{appendix:convergence}
This appendix investigates whether our primary results are sensitive to our choice of gravitational softening length and time stepping. A worry is that the axion cosmology is so much different than the standard cosmology that simulations may require much different accuracy parameters.  We find that this is not the case.  

To test convergence, we have run a reference simulation that uses the same box size, gravitational softening parameter and time-stepping parameter as the simulations in Sec.\ref{sec:simulation}. We compare it to a simulation with half the gravitational softening length and one with twice as small time steps.  All simulations considered in this appendix take $N=512^3$ particles and are run in a box size of 50 pc/h.

In particular, the reference gravitational softening length is equal to 1/30 in units of the mean separation of N-body particles. (The mean separation can be estimated as $L/N_{\rm grid}$, where $L$ is the box size and $N_{\rm grid}$ is the cube root of the number of particles.) In the reference run, the time stepping parameter $\eta$ is taken to be 0.02, which is the default time-stepping parameter in MP-Gadget.  The time-stepping criterion \cite{2003MNRAS.338...14P} used in MP-Gadget:
\begin{equation}
    \Delta t =\sqrt{2\eta\epsilon/a},
\end{equation}
where $\epsilon$ is the gravitational softening length, $\eta =\textit{ErrTolIntAccuracy}$ is a dimensionless parameter and $a$ is the local acceleration.  The error of MP-Gadget's leap-frog integration scheme scales quadratically in $\Delta t$. 
 Both the softening and time stepping parameters are set to half their reference values in our two additional simulations to test for convergence.

\begin{figure*}
\includegraphics[width= 0.9\textwidth]{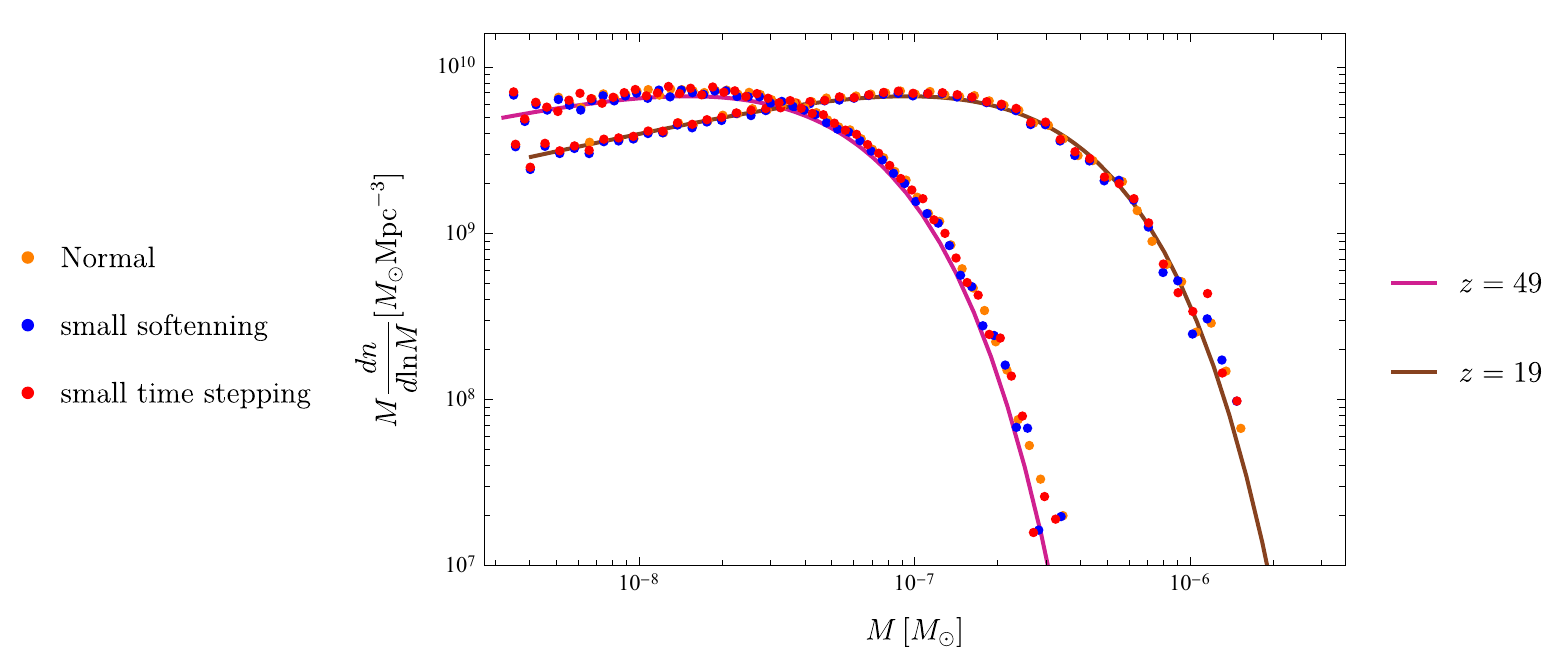}
\caption{Simulated FOF mass function at $z=49$ and $z=19$. Data points with different colors are from simulations run with different force softening and timestepping parameters, in addition to the fiducial simulation.  See the appendix for more information on the different parameter values. The solid curves are the semi-analytic prediction discussed in Appendix~\ref{appendix:Fitting}. The simulated mass functions all agree with each other, suggesting that the halo mass function is converged in softening and time stepping.
}
\label{fig:MF_comparison}
\end{figure*}

\begin{figure*}
\includegraphics[width= 1\textwidth]{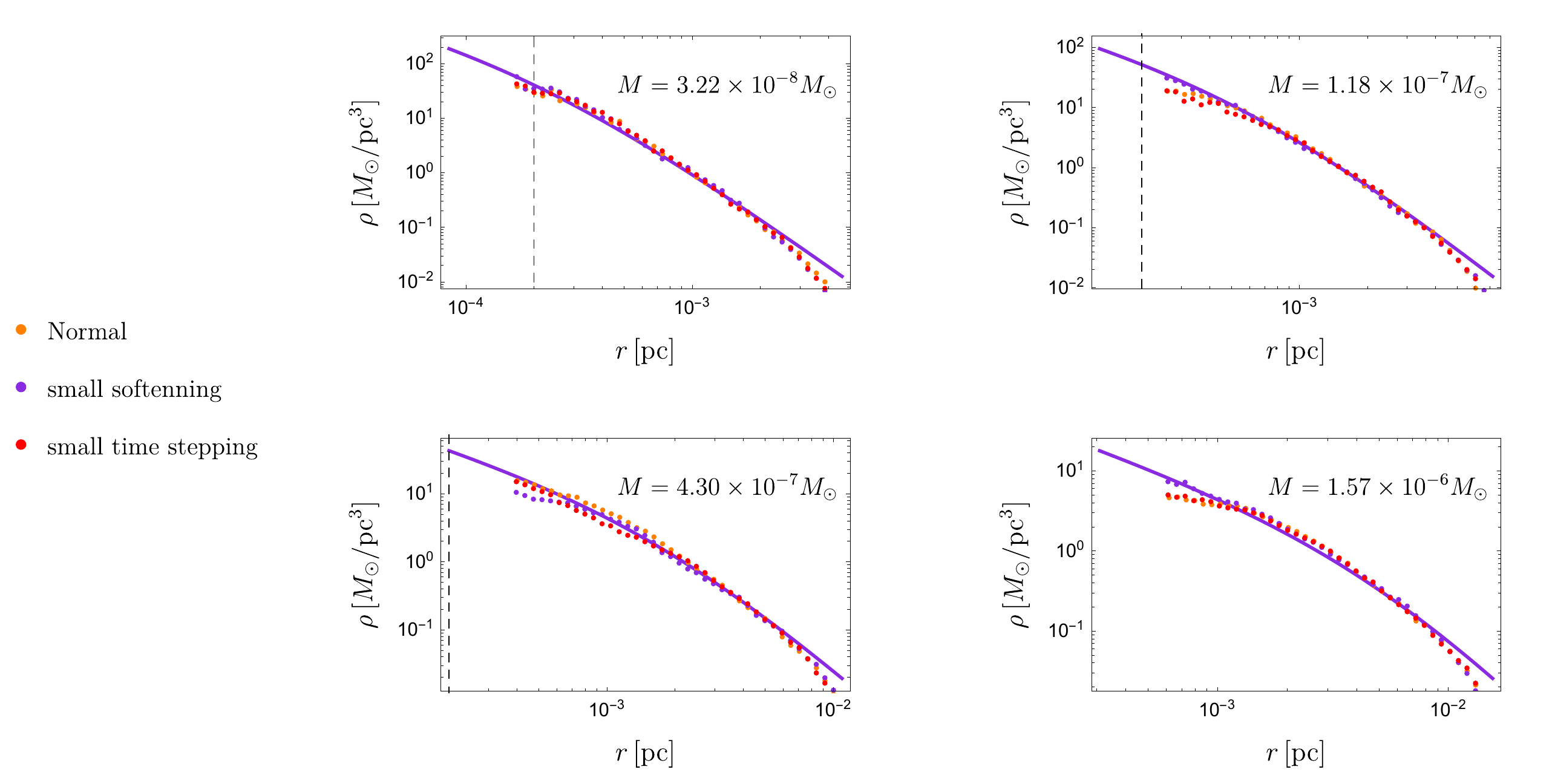}
\caption{Density profiles at $z=19$ for halos with different masses. These plots are obtained by averaging 20 halos around the specified mass. The orange, purple and red data points are from the reference simulation, the simulation with a smaller gravitational softening length, and the simulation with a smaller time-stepping parameter, respectively. The solid curves are the NFW halo profile with scale radius given by Eq.~\ref{eq:scale_radius}. The vertical dashed line indicates the fiducial gravitational softening length in our simulation.
}
\label{fig:dens_pro_comparison}
\end{figure*}

We compare the mass function  and the density profile in the reference run and the convergence-testing runs, as shown in Fig.~\ref{fig:MF_comparison} and Fig.~\ref{fig:dens_pro_comparison}.  We plot the mass functions at $z=19$ and $z=49$, as shown in Fig.~\ref{fig:MF_comparison} and the mass functions of all the simulations agree well and with our semi-analytic model.

The $z=19$ density profiles in  Figure~\ref{fig:dens_pro_comparison} are obtained by averaging 20 halos around the specified mass (points) as well as the NFW halo profile with scale radius given by Eq.~\ref{eq:scale_radius} (solid curves). We show density profiles of halos at four different masses, and they all agree well with the NFW profile. There are some differences at smaller radii, although the differences do not notably go in one direction or the other.  We suspect this owes to slight differences in the properties of halos being used in the average.
Overall, the halo mass function and density profile appear to be convergence.  
In Fig. \ref{fig:mass_function}, we also compared the mass function with simulation that has a different box size $10 \,\rm pc$ and particle number $512^3$, which shows good convergence as well.
We also note that Fig. \ref{fig:dens_pro_comparison} is showing the lowest redshift in our simulation, where the comoving softening length subtends the largest physical scale.

\section{Equivalence of Disruption Timescales}\label{appendix:disruption}
When the encounter velocity $V$ is much greater than the minihalo's internal dispersion, the impulse approximation may be used. For minihalos, all encounters are in this regime. When the impact parameter $b \gg r_s$, the characteristic radius of the minihalo, the distant-tide  approximation is valid, allowing the perturbing potential to be expanded and higher order terms dropped. This introduces error in the low $b$ regime which generally must be investigated with simulations. For minihalos, \citet{Green_2007} investigated this and found that a good approximation to the energy imparted from a high speed encounter with stars is:
\begin{align} \label{eq:enc_energy_complete}
    \Delta E &= \frac{4\alpha^2}{3} \frac{G^2M_p^2 M r_{\rm vir}^2}{V^2b^4} \quad& \mathrm{(b > b_s)}; \\
    \Delta E &= 3\beta^2 \frac{G^2M_p^2 M }{V^2r_{\rm vir}^2} \quad& \mathrm{(b < b_s)}.
\end{align}
where $b_s \equiv(4\alpha^2/9\beta^2)^{1/4}r_{\rm vir}$.  As discussed in \S~\ref{sec:disruption}, we find that for all but the largest minihalos in cluster environments, the $b>b_s$ expression applies. Here we show the derivation of the  $R_H=R_c$ equivalence under this assumption.

We assume the perturber and subject have velocities drawn from Maxwellian distributions, so that the relative encounter velocities are also Maxwellian, with dispersion $\sigma_{\rm rel}$. The average rate of encounters with perturbers of number density $n_p$, encounter speed $V$, and impact parameter $b$ is \cite{Binney2008}:
\begin{equation}\label{eq:encounter_rate}
    \dot{C} = \frac{2 \sqrt{2 \pi} n_p}{\sigma_{\rm rel}^3} \exp{\bigg\{-\frac{V^2}{2\sigma_{\rm rel}^2}\bigg\}}V^3 \mathrm{d}V  b \mathrm{d}b.
\end{equation}
The catastrophic encounter rate is then obtained from integrating Eq.~\ref{eq:encounter_rate} over all encounter velocities and impact parameters from 0 to $b_c$:
\begin{eqnarray}
    R_c &\equiv& \frac{2\sqrt{2\pi}n_p}{\sigma_{\rm rel}^3} \int_0^{\infty}  V^3 \mathrm{d} V \exp{\bigg\{-\frac{V^2}{2\sigma_{\rm rel}^2}\bigg\}} \int_0^{b_c} b \, \mathrm{d} b  \nonumber \\
    &=& \sqrt{\frac{8G\alpha^2}{3f_c M}}\pi \rho_p r_{\rm vir}^{3/2},
\end{eqnarray}
where the maximum catastrophic impact parameter $b_c$ is found by setting the ratio of the energy impacted over the binding energy to unity:
\begin{equation}
    b_c  = \bigg( \frac{8\alpha^2G M_p^2 r_{\rm vir}^3}{3 f_c M V^2} \bigg)^{1/4}.
\end{equation}

The remaining encounters, $b>b_c$, are in the diffusive regime.  Each encounter creates some small $|\Delta \boldsymbol v| \ll \boldsymbol v$ impulse on the particles, changing the energy per unit mass by $\boldsymbol v \cdot \Delta \boldsymbol v + |\frac{1}{2}\Delta \boldsymbol v|^2 $. The first term is much larger, but we assume there is no preferred angle and so it cancels out, leaving the heating term. The average rate of energy added from non-catastrophic encounters is $\dot{E} = \dot{C}  \left\langle \Delta E \right\rangle$. We use Eq.~\ref{eq:encounter_rate} and the high $b$ expression for $\Delta E$ (Eq.~\ref{eq:enc_energy}) and integrate over all encounter velocities and from $b_c$ upwards:
\begin{equation}
\begin{split}
      \dot{E} &= \dot{C} \left\langle \Delta E \right\rangle = \frac{8\alpha^2}{3}\sqrt{2\pi} \frac{G^2 M M_p^2 n_p r_{\rm vir}^2}{\sigma_{\rm rel}^3} \\
      & \int_{0}^{\infty} V \mathrm{d} V \exp{\bigg\{-\frac{V^2}{2\sigma_{rel}^2}\bigg\}} \int_{b_c}^{\infty} \frac{\mathrm{d} b}{b^3}, \\
      &= \sqrt{\frac{\alpha^2|E_b|M}{3}} 2\pi  G \rho_p r_{\rm vir}.
  \end{split}
\end{equation}
Finally, we obtain the fractional heating rate by dividing by the binding energy:
\begin{equation}
    \begin{split}
    R_H \equiv \frac{\dot{E}}{|E_b|} = \sqrt{\frac{8G\alpha^2}{3f_c M}}\pi \rho_p r_{\rm vir}^{3/2} =  R_c.
    \end{split}
\end{equation}
We arrive at the surprising conclusion that, so long the use of the high $b$ expression for $\Delta E$ is justified, the expected time for one catastrophic encounter is the same as the average time taken for cumulative encounters to impart energy equal to the binding energy.  

Previous heating estimates in the literature using the impulsive encounter framework often use a low $b$ cutoff of $\sim r_{\rm vir}$, which introduces dependencies on $\sigma_{\rm rel}$ and the mass distribution of perturbers. This is justified when $b\sim r_{\rm vir}$ encounters are generally not catastrophic, and when penetrative encounters should impart negligible energy. However, when the catastrophic regime exists, the heating calculation should begin at $b_c$, where the catastrophic encounters ended, to correctly count all encounter impact parameters. This leads to the equivalence of the disruption timescales.  


\bibliographystyle{apsrev4-2}
\bibliography{axionhalo}
\end{document}